\documentclass[12pt]{article}

\usepackage{amsmath} \usepackage{amsfonts}

 \usepackage{pictexwd,dcpic}

\newcommand\bm[1]{\mbox{\boldmath$#1$}}

\def\grad{\mathrm{\scriptscriptstyle grad}}
\def\gradL{\mathrm{grad } \,}

\def\supIe{\Sigma^{+}_{\ee}}
\def\mmIe{\M^{+}_{\ee}}

\def\supIo{\Sigma^{+}_{0}}
\def\mmIo{\M^{+}_{0}}

\def\X{X}
\def\W{W}

\def\qper{{h^{\prime}}}
\def\qperper{{h^{\prime\prime}}}
\def\kappaper{{\kappa^{\prime}}}
\def\kappaperper{{\kappa^{\prime\prime}}}

\def\Kper{{K^{\prime}}}
\def\Kpergauge{{K^{\prime}_{g}}}

\def\Kperper{{K^{\prime\prime}}}
\def\Kperpergauge{{K^{\prime\prime}_{g}}}

\def\Kpernornor{{Y^{\prime}}}
\def\Kperpernornor{Y^{\prime\prime}}
\def\Kpertan{{\Kper^{\,t}}}
\def\Kperpertan{{\Kperper^t}}
\def\Kpernortan{{\tau^{\prime}}}
\def\Kperpernortan{{\tau^{\prime\prime}}}

\def\ee{\epsilon}
\def\gee{g_{\ee}}
\def\aln{\alpha_1 \cdots \alpha_m}
\def\almun{\alpha_1 \cdots \mu \cdots \alpha_m}

\def\mun{\mu_1 \cdots \mu_m}

\def\murhon{\mu_1 \cdots \rho \cdots \mu_m}

\def\U{\cal U}

\def\P{P}
\def\AA{{\cal A}}

\def\none{{n_1}}
\def\ntwo{{n_2}}
\def\mone{{m_1}}
\def\mtwo{{m_2}}

\def\gzero{g}
\def\gone{\overline{g}}

\def\nablazero{\nabla}
\def\nablaone{\overline{\nabla}}

\def\Qone{Q_1}
\def\Qtwo{Q_2}

\def\T{{{\cal T}}}
\def\Tone{{T_1}}
\def\Ttwo{{T_2}}

\def\Cone{{C_1}}
\def\Done{{D_1}}

\def\Z{{Z}}
\def\Zone{{Z_1}}
\def\Ztwo{{Z_2}}

\def\Sone{{S_1}}
\def\Stwo{{S_2}}

\def\uone{{u_1}}
\def\utwo{{u_2}}

\newcommand{\Sper}{{{\cal S}^{\prime}}}
\newcommand{\Sperper}{{{\cal S}^{\prime\prime}}}

\newcommand{\zee}{{\vec{Z}_{\ee}}}

\newcommand{\be}{\begin{eqnarray}}
\newcommand{\en}{\end{eqnarray}}
\newcommand{\bestar}{\begin{eqnarray*}}
\newcommand{\enstar}{\end{eqnarray*}}     
\newcommand{\M}{{\cal    M}}
\newcommand{\Mp}{{{\cal   M}^{+}}}   
\newcommand{\Mm}{{{\cal  M}^{-}}}
\newcommand{\Mpm}{{{\cal  M}^{\pm}}}  
\newcommand{\Mt}{{{\cal M}^{t}}}
\newcommand{\gp}{{g^{+}}} 
\newcommand{\gm}{{g^{-}}}
\newcommand{\gt}{{g^{t}}} 
\newcommand{\gpm}{{g^{\pm}}}
\newcommand{\N}{{\cal   N}} 
\newcommand{\ji}{{{\mathfrak   S}  (\M)}}
\newcommand{\nep}{{\nabla^{\ee}}}

\newcommand{\C}{{\cal       C}}      
\newcommand{\calT}{{\cal     T}}

\def\Journal#1#2#3#4#5#6{#1, ``#2'', {#3} {\bf #5}, #6 (#4).}

\def\CQG{\em Class. Quantum Grav.}

\def\PRD{\em Phys. Rev. D }

\def\MNRAS{\em Mon. Not. Roy. Astr. Soc.}

\def\CMP{\em Commun. Math. Phys.}

\def\PRL{\em Phys. Rev. Lett.}

\def\AJ{\em Astrophysical Journal}

\def\PLB{\em Phys. Lett. A}
\def\PTPS{\em Prog. Theor. Phys. Supp.}
\def\NPB{\em Nuclear Phys. B}
\def\IJMPA{\em Int J. Mod. Phys A}

\def\LRR{\em Living Reviews in Relativity}

\newtheorem{lemma}{Lemma} 
\newtheorem{theorem}{Theorem}

\newtheorem{proposition}{Proposition}

\begin{document}

\bibliographystyle{amsplain}

\title{First and  Second   Order  Perturbations of Hypersurfaces}
\author{Marc    Mars
\\ Facultad de  Ciencias, Universidad de Salamanca \\
Plaza  de   la  Merced  s/n,  Salamanca  37008,   Spain.   \\  e-mail:
marc@usal.es}
\maketitle
\begin{abstract}
In this paper we find the first and second order perturbations of the induced metric and the extrinsic
curvature of a non-degenerate hypersurface $\Sigma$ in a spacetime $(\M,g)$, when the metric $g$
is perturbed arbitrarily to second order and the hypersurface itself is
allowed to change perturbatively (i.e. to move within spacetime) also to second order. The results are
fully general and hold in arbitrary dimensions and signature. An application of these results
for the perturbed matching theory between spacetimes is presented.
\end{abstract}


\newpage

\section{Introduction}

The aim of this paper is to analyze how the induced metric and extrinsic
curvature of a hypersurface change when the spacetime metric
is perturbed to second order and the 
hypersurface itself  is deformed perturbatively to second order.
The motivation to
carry out this calculation is twofold. Firstly, to study
the matching conditions between two spacetimes when the metric is perturbed arbitrarily
to second order. Secondly, to study the dynamics of first and second order
perturbations of an $(n-1)$-brane and its backreaction on the bulk.
These two problems involve in an essential way knowing the first and
second order perturbations of the induced metric and extrinsic curvature of the hypersurface.

A natural question that should be addressed to start off is whether going to second
order in perturbation theory is necessary and interesting. Taking for granted that
perturbation theory is useful and powerful for a wealth problems, the point is why
second order. First of all, our present understanding of linear perturbations (methods
involved, subtleties, applications. etc.) has reached a level of maturity
that allows us to go to second order as a natural 
next step. More importantly, there are many situations where linear theory
is not accurate enough and second order non-linear effects have to be considered. 
One  fundamental example is studying inhomogeneities in our Universe 
(see \cite{Carbone05}, \cite{Kolb05}, \cite{Malik04}, \cite{Mena02}, \cite{Matarresse98}, and references therein)
and particularly structure formation (e.g. \cite{Singh02},\cite{Watts01}). Another example
is anisotropies in the cosmic microwave 
background, where  the present and planned sensitives of the detectors
is high enough so that second order effects can already be detected (or ruled out) (see
\cite{Bartolo04.1}, \cite{Bartolo04.2}, \cite{Acquaviva03}
for some recent references).

Besides cosmology, second
order perturbations are also fundamental in slowly rotating
stars. Since the seminal work of Hartle \cite{Hartle67} we know that stationary and axially
symmetric (rotating) perturbations of static stars have a linear component
which sets the star into rotation without modifying the shape of its boundary. Its effect to
second order is to modify the shape of the star as well as to modify  
the rest of the metric components. 
Dynamical (non-stationary)
perturbations of stars (both static or slowly rotating) also require going to second order
(see \cite{Stergioulas} and references therein). 
Second order perturbations have also been applied to back holes, specially to study the close
limit in the collision of back holes (\cite{Gleiser96}, \cite{Gleiser00} and therein references).
When compared with numerical work, the results of these calculations show
that the range of validity of the perturbative regime is much broader than expected.

In brane scenario, perturbations are also relevant.
One important aspect has to do with stability of
branes, for which a proper understanding of how the intrinsic
geometry of the branes behaves under perturbations is required. 
Small (first order)
perturbations of a worldsheet evolving in a fixed flat background were studied in
\cite{Garriga93}. The generalization to curved background (still fixed) was done in
\cite{Carter93}, \cite{Guven93}. Including the backreaction of the bulk (i.e.
first order perturbations of a brane in a perturbed bulk) was analyzed in
\cite{Battye95}, where in particular the linear perturbation of the induced metric and
second fundamental forms of an arbitrary (non-null) brane were calculated. The calculations
were carried out using exclusively spacetime tensors.
This approach is very efficient because there is no need to use
two different
sets of indices (spacetime indices and intrinsic indices on the brane). 
Linear perturbations of branes have also been studied in the context of 
cosmological perturbations on the brane (see \cite{reviews} for reviews). 
Different approaches have been considered:
the most widely used involves calculations in explicit coordinate systems (see
\cite{MalikRodriguez-MartinezLanglois} and references therein), another approach uses a covariant formalism
\cite{Maartens02} and a third method is the doubly gauge
invariant formalism developed by Mukohyama \cite{Mukohyama00}. The latter is geometrically very clear
and has the advantage
that the gauge freedom in the bulk and on the brane are neatly separated. As far as I know \cite{Mukohyama00}
and \cite{Battye95} are the only papers where general first order perturbations of an arbitrary non-null
hypersurface in an arbitrary spacetime are studied. 

Second order calculations have also been considered 
for some highly symmetric bulks and branes 
\cite{KudohTanaka01},\cite{Kudoh02},\cite{KudohTanaka02},\cite{KudohTanaka03}.
So far, the only completely general calculation to second order
can be found in \cite{Battye00} where the Lagrangian density of the Dirac-Goto-Nambu brane
was calculated when both the bulk and of
the worldsheet were perturbed to second order. The main motivation in \cite{Battye00}  was to find a
quadratic Lagrangian for the first order perturbation equations. 
The results in the present paper 
generalize those in \cite{Battye00} and extend the calculation 
of the induced volume form (i.e. the Lagrangian) to the full first 
and second fundamental forms. However, 
only codimension one branes are considered here, unlike \cite{Battye00} where any codimension was allowed.

Gauge invariance is a subtle point in any geometrical perturbation theory. 
In the case of hypersurfaces the complexity increases due to the presence of a moving boundary, and this has led
to some confusion in the literature. Going to second order complicates matters both conceptually and 
in the calculations. Thus, it becomes necessary to
state clearly the theoretical framework defining the perturbations.
In this paper I will use a geometrical method
that, in some sense, combines the approaches of \cite{Battye95} and  \cite{Mukohyama00}. 
As in \cite{Mukohyama00}, the hypersurface
and its perturbations will be defined by embeddings of an abstract manifold into the spacetime (thus splitting
the intrinsic objects from the ambient spacetime objects in a clear way). This allows us to separate the two gauge freedom
sources neatly. For the explicit calculation of perturbations, it is much more economic to 
use the spacetime formalism, as in \cite{Battye95}. The result is finally written
down in terms of intrinsic tensors on the unperturbed hypersurface, where all objects are naturally defined.
The calculations
leading to second order perturbations are difficult.
If a correct approach is not taken, the expressions easily become very large and 
unmanageable. Thus, I will spend some effort explaining how
the calculations are performed. A basic ingredient is a
Lemma which relates perturbations of intrinsic objects to a hypersurface
and perturbations of suitable objects in the ambient space. This Lemma will be used not only to calculate the perturbations
within the hypersurface but also to analyze the second order gauge freedom. This approach to the gauge transformations
is complementary to that in \cite{Bruni97}, {\cite{Sonego98} where the so-called
knight diffeomorphisms were used.

The calculations presented in this paper have many potential applications. One which I consider very
relevant is the matching of spacetimes. 
Constructing spacetimes from the junction of two regions across their
boundary has been a very useful tool in gravitational theory. The set
of conditions ensuring that two spacetime regions can be joined 
are well understood (see e.g. \cite{MarsSenovilla93} for a detailed account).
The role of the discontinuity of the second fundamental
form and its relationship with distributional parts in the
energy-momentum tensor were clarified by Israel and have become known
as Israel matching conditions \cite{Israel}. Since perturbation theory has been useful for 
many problems, it is not surprising that perturbing spacetimes 
constructed from matching two regions is also of interest. Obvious examples
are perturbed stars, voids in the universe (as in the Einstein-Straus cheese model \cite{ES}),
shells of matter, impulsive waves, etc. In this paper I will present, as a by product, 
the first and second order perturbed matching conditions for non-null hypersurfaces.

The paper is intended to be self-contained, so that all the subtleties and difficulties
of second order calculations on hypersurfaces can be properly understood. I will, of course, give credit
to previous results
whenever necessary. The paper is organized
as follows. In section 2, a brief summary of perturbation theory is given. This fixes our
framework and notation. In section 3, the Lemma mentioned above  will be stated and proven.
This result will be used throughout the paper.
A consequence of this Lemma is that second order
perturbations of
hypersurfaces can be described by two vector fields defined on the unperturbed hypersurface. Their
explicit form in an arbitrary coordinate system is discussed in Section 4. Section 5 deals with first
order perturbations of the fundamental forms. Here the results of \cite{Battye95} and 
\cite{Mukohyama00} mentioned above will be  recovered.
In this section some useful Lemmas to carry
out the second order calculations are presented. The result of the second order perturbations is
stated in Section 6, leaving the details of the proof to Appendices A and B.
Section 7 discusses first and second order
gauge transformations. For hypersurfaces there are two types of gauge freedom, namely the one coming from the ambient spacetime
and another one intrinsic to the hypersurface. Both are discussed in this  section. Section 8 applies the previous results 
to the perturbed matching conditions between two spacetimes. A theorem giving the necessary and sufficient conditions
for second order perturbations to match across a matching hypersurface in the background is presented. This theorem
can be potentially applied to many situations. One case that has already been investigated
involves first and second order stationary and axially symmetric perturbations of spherical stars \cite{Malcolm}.
The two Appendices contain
the main steps in the calculations of the second order perturbations. 
The reader who is not interested in detailed calculations may skip the Appendixes and concentrate on the main text.

\section{Summary of  perturbation theory}

Perturbation theory   deals    with   one   parameter   families   of
spacetimes\footnote{ A $C^{m}$  spacetime is a Hausdorff,
connected  $C^{m+1}$   manifold  of  dimension  $n$   endowed  with  a
Lorentzian   metric  of   class   $C^m$.   If we are considering a
manifold-with-boundary then the boundary $\partial \M$ is also assumed
to  be  $C^{m+1}$. 
Our signature and sign conventions for the 
Riemann and Ricci
tensors  follow \cite{Wald}.}  $(\M_\ee,\hat{g}_\ee)$ 
and their first and higher order variations around one 
element of the family, say
$(\M_{0},g_{0})$. In order to take derivatives with respect to $\ee$ we
need some fixed set of points, i.e. we need all
$\M_{\ee}$ to be diffeomorphic to $\M_{0}$. Through these diffeomorphisms we can define a
one-parameter family of metrics $g_{\ee}$ on $\M_0$ associated to $\hat{g}_{\ee}$.
We denote $\M_0$ simply as $\M$.
Since we want to take second variations of extrinsic curvatures we take $\M$ to be 
$C^4$, as a manifold. We  thus consider a differentiable family of Lorentzian metrics on $\M$,
i.e. a $C^2$ map
\begin{eqnarray*}
T \,  \, : \,  \, I &  \longrightarrow & \ji  \nonumber\\ \, \,  \, \,
\ee & \longrightarrow & g_\ee
\end{eqnarray*}
where $\ji$  denotes the set of $C^3$  symmetric, non-degenerate, two-index tensor
fields in $\M$ ($\ji$ can be endowed with a natural differential structure, $T$ is $C^2$ with respect
to this structure).

We denote by  $\nep$ the Levi-Civita covariant derivative of
the metric $g_\ee$. Given   two
arbitrary   metrics  $\gzero$  and   $\gone$,  the corresponding  covariant
derivatives  $\nablazero$ and  $\nablaone$  are well-known \cite{Wald} to be related
by
\begin{eqnarray}
\nablaone_{\mu} T^{\alpha}_{\beta}  = \nablazero_{\mu} T^{\alpha}_{\beta}
+   \C^{\nu}_{\beta\mu}    T_{\nu}^{\alpha}   -   \C^{\alpha}_{\nu\mu}
T_{\beta}^{\nu},
\label{relcovs}
\end{eqnarray}
where     $\C^{\alpha}_{\beta\gamma}    =     (1/2)    \gone^{\alpha\mu}
(\nablazero_{\beta}  \gone_{\mu\gamma} + \nablazero_{\gamma}  \gone_{\mu\beta} -
\nablazero_{\mu}  \gone_{\beta\gamma}  )$.   Similar expressions  hold  for
tensors with any  number of indices. 
Substituting $\gzero$  by $g_{\ee=0}$ and $\gone$  by $g_{\ee}$, we
can take $\ee$-derivatives in (\ref{relcovs}) at $\ee=0$ to get
\begin{eqnarray}
\left  . \frac{\left  (  d \nabla^{\ee}_{\mu}  T^{\alpha}_ {\beta}
 \right  ) }{d  \ee}\right |_{\ee=0}  = \Sper^{\nu}_{\beta\mu}
 T_{\nu}^{\alpha} - \Sper^{\alpha}_{\nu\mu} T_{\beta}^{\nu},
\label{FirstT}
\end{eqnarray}
where we have defined
\begin{eqnarray}
\Kper_{\alpha\beta}
\equiv \left .  \frac{d g_{\ee \, \alpha\beta}}{d \ee} \right
|_{\ee=0}, 
\quad  
\Sper^{\alpha}_{\beta\gamma}  \equiv \frac{1}{2} \left  ( \nabla_{\beta}
\Kper^{\alpha}_{\,\gamma}  +  \nabla_{\gamma}  \Kper^{\alpha}_{\,\,\beta}  -
\nabla^{\alpha}  \Kper_{\beta\gamma}  \right  ).
\label{Sper}
\end{eqnarray}
Here and  in the following we  set $g_{0} \rightarrow g$  and  
$\nabla^0 \rightarrow \nabla$.
The tensor  $\Kper$ is the
first   order   perturbation    of  $g$   along   the   family
$g_{\ee}$. We will simply call it {\it first order perturbed
metric}. Similar expressions as (\ref{FirstT}) hold for tensors with any number of indices. 
The second  derivatives   of  (\ref{relcovs})  gives
\begin{eqnarray*}
\left .  \frac{\left  ( d^2 \nabla^{\ee}_{\mu} T^{\alpha}_{\beta}
 \right ) }{d \ee^2}\right |_{\ee=0} = \left ( \Sperper^{\nu}_{\beta\mu}
+ 2 \Kper^{\nu}_{\,\, \delta} \Sper^{\delta}_{\beta\mu} \right ) T_{\nu}^{\alpha}  
-  \left ( \Sperper^{\alpha}_{\nu\mu}  
+ 2 \Kper^{\alpha}_{\,\, \delta} \Sper^{\delta}_{\nu\mu} \right ) T_{\beta}^{\nu},  
\end{eqnarray*}
where 
$\Kperper_{\alpha\beta}
\equiv    \frac{d^2    g_{\ee \, \alpha\beta}}{d    \ee^2}
|_{\ee=0}$ 
is the second order 
perturbed metric  and
\begin{eqnarray}
\Sperper^{\alpha}_{\beta\gamma} \equiv  \frac{1}{2} \left ( \nabla_{\beta}
\Kperper^{\alpha}_{\,\gamma}  + \nabla_{\gamma}  \Kperper^{\alpha}_{\,\,\beta} -
\nabla^{\alpha} \Kperper_{\beta\gamma} \right ). \label{Sperper}
\end{eqnarray}

\section{A useful Lemma}
\label{UsefulLemma}

We want to calculate how the first and second fundamental forms of a hypersurface
change when the ambient metric and the hypersurface are perturbed. Thus, we
consider a one-parameter family of hypersurfaces $\Sigma_{\ee}$ of $\M$. 
As before, in order to define variations we need a fixed  set of points, so
we assume all $\Sigma_{\ee}$ to be diffeomorphic to each other.
We allow the hypersurfaces to change both as a subset of points of $\M$ and also in the way
we coordinate them for different $\ee$. Since the fundamental forms are
pull-backs of tensors on $\M$, their dependence on $\ee$ arise because of three facts:
(i) because the ambient metric depends on $\ee$,
(ii) because the hypersurface $\Sigma_{\ee}$ considered as a subset of $\M$ changes with $\ee$ and
(iii) because the way in which we coordinate $\Sigma_{\ee}$ is allowed to depend on $\ee$
(even if the hypersurface as a set of points remains unchanged). Points (ii) and (iii) 
can be treated together by viewing the hypersurfaces $\Sigma_{\ee}$ as embedded hypersurfaces,
i.e. as the images of a family of embeddings $\Phi_{\ee}: \Sigma \rightarrow \M$,
where $\Sigma$ is a copy of any of the $\Sigma_{\ee}$, say $\Sigma_0$. It is useful to view
$\Sigma$ as an abstract manifold
detached from the spacetime so that one knows clearly where the different objects are defined.
Thus, we shall distinguish between $\Sigma$ as an $(n-1)$-dimensional manifold and $\Sigma_0 = \Phi_0 (\Sigma)$, which
is the hypersurface in $\M$.

The fact that the fundamental forms on $\Sigma$ depend on $\ee$ from several sources
and that we want to do the calculation up to second order 
makes it very important to use a method as covariant as possible
and use coordinates only when absolutely necessary. Moreover, it is very convenient to
work on the ambient manifold as much as possible and perform the pull-back only at the very end (in agreement with
\cite{Battye95}). The
alternative of calculating the derivative directly on $\Sigma$ is, of course, possible but much more difficult.
In this section I present a Lemma which 
shows how derivatives of geometric tensors on an embedded submanifold (or arbitrary codimension,
including codimension $0$) 
can be calculated from derivatives performed purely on the ambient manifold. This result will be crucial
for the calculations in the following section. Throughout this section all differentiable objects
are $C^3$ unless otherwise specified.

Thus, let $\N$ and $\M$ be two differentiable manifolds and
let  $\chi_{\ee}  : \N  \rightarrow  \M$  be  a family  of  differentiable maps. 
Let  us  consider  a $C^2$  family  of  
covariant tensor fields $T_{\ee}$ on $\M$. We can pull-back this family to $\N$
and define a one-parameter  family of tensors $\calT_{\ee}$ on $\N$.
We are interested  in determining the first and  second derivatives of
$\calT_{\ee}$  with  respect to  $\ee$.  Using directly the definition of derivative,
\begin{eqnarray}
\frac{d \calT_{\ee}}{d \ee}  = \mathop{\lim} \limits_{h \mathop
\to 0 } \frac{ \chi^{\star}_{\ee+h} \left ( T_{\ee+h} \right
)  - \chi^{\star}_{\ee}  \left  ( T_{\ee}  \right  ) }{h}  =
\mathop{\lim}     \limits_{h     \mathop     \to    0     }     \frac{
\chi^{\star}_{\ee+h}   \left   (   T_{\ee+h}  \right   )   -
\chi^{\star}_{\ee}  \left  (   T_{\ee+h}  \right  )  }{h}  +
\chi^{\star}_{\ee} \left ( \mathop{\lim} \limits_{h \mathop \to 0
} \frac{ T_{\ee+h} - T_{\ee}}{h} \right ),
\label{DerivativeTep}
\end{eqnarray}
where  we  have   added  and  subtracted  $\chi^{\star}_{\ee}  (
T_{\ee+h} )$ in  the numerator and we have  used the linearity of
$\chi^{\star}_{\ee}$.  The  second  term is
the  pull-back of the derivative  of $T_{\ee}$. The first term cannot be written directly
in a simple form because
there is no a priori relationship  between
$\chi^{\star}_{\ee+h}$ and $\chi^{\star}_{\ee}$ (contrarily
to  what happens for instance in a one-parameter group  of
diffeomorphisms). Assume now that $\chi_{\ee}$ are
embeddings. Then, there   exists  a  set  of
diffeomorphisms $\Psi_h^{\ee} : \M \rightarrow \M$ of the ambient space such that the
diagram
\begin{equation}
\begindc{\commdiag} \obj(1,3){$\N$}[A] \obj(1,1){$\M$}[B]
\obj(3,1){$\M$}[C] \mor(1,3)(1,1){$\chi_{\ee}$}
\mor(1,3)(3,1){$\chi_{h+\ee}$}
\mor(1,1)(3,1){$\Psi_{h}^{\ee}$} \enddc
\label{DiagramComm}
\end{equation}
is commutative. We moreover fix $\Psi_{0}^{\ee} = \mathbb{I}_{\M}$ (the identity on $\M$) for all $\ee$.
Geometrically, $\Psi_{h}^{\ee}$ transforms the point $\chi_{\ee}(p)$ into the point $\chi_{\ee+h}(p)$ for all
$p \in \N$. Notice that $\Psi_{h}^{\ee}$ is fixed only on 
$\chi_{\ee} (\N)$ and therefore it is non-unique in general.
However, when $\N$ has the same dimension as $\M$ and $\chi_{\ee}$
are diffeomorphisms, then
$\Psi_{h}^{\ee}$ is unique and given by $\chi_{\ee+h} \circ \chi^{-1}_{\ee}$.

The following Lemma gives explicit expressions for
the first and second $\ee$-derivatives of $\T_{\ee}$ in terms of objects defined in the ambient
space $\M$.
\begin{lemma}
\label{FirstSecondOrder}
Let $T_{\ee}$ be a  $C^2$ one-parameter family of covariant
tensor fields on $\M$, $\chi_{\ee}: \N \rightarrow \M$ a $C^2$
family of embeddings and define  $\calT_{\ee} =
\chi_{\ee}^{\star} ( T_{\ee} )$. Then
\begin{eqnarray}
\frac{d \calT_{\ee}}{d  \ee} = \chi_{\ee}^{\star}  \left (
\pounds_{\vec{\Z}_{\ee}}  T_{\ee}  + \frac{d  T_{\ee}}{d
\ee}  \right  ),  \label{depsilon} \\ 
\frac{d^2  \calT_{\ee}}{d  \ee^2}  =
\chi_{\ee}^{\star}    \left    (    \pounds_{\vec{\W}_{\ee}}
T_{\ee}               +              \pounds_{\vec{\Z}_{\ee}}
\pounds_{\vec{\Z}_{\ee}}          T_{\ee}         +         2
\pounds_{\vec{\Z}_{\ee}} \left ( \frac{d T_{\ee}}{d \ee}
\right ) + \frac{d^2 T_{\ee}}{d \ee^2} \right ), \label{d^2epsilon}
\end{eqnarray}
where
\begin{eqnarray}
\vec{\Z}_{\ee}   =  \left .  \frac{\partial  \Psi_h^{\ee}}{
\partial  h}  \right  |_{h=0},  \hspace{2cm}  
\vec{\W}_{\ee}  = \frac{d  \vec{\Z}_{\ee}}{d  \ee} 
\label{Zs}
\end{eqnarray}
and $\Psi_{h}^{\ee}$ is any set of diffeomorphisms of $\M$ which
makes the diagram (\ref{DiagramComm}) commutative.
\end{lemma}

{\it Proof:} The commutativity of the diagram
({\ref{DiagramComm}) implies that the first term in (\ref{DerivativeTep})
can be written as
\begin{eqnarray*}
\mathop{\lim} \limits_{h \mathop  \to 0 } \frac{\chi^{\star}_{\ee
+ h} \left (T_{\ee +  h} \right ) - \chi^{\star}_{\ee} \left
(T_{\ee+h}  \right  )  }{h}  =  \chi^{\star}_{\ee}  \left  (
\mathop{\lim}   \limits_{h  \mathop   \to  0   }  \frac{\Psi^{\ee
\star}_{h}  \left (T_{\ee  + h}  \right )  -  T_{\ee+h} }{h}
\right    )    =    \chi^{\star}_{\ee}    \left    (    \pounds_{
\vec{\Z}_{\ee} } T_{\ee} \right ),
\end{eqnarray*}
where the definition of Lie derivative has been used. This proves 
(\ref{depsilon}). For 
the second derivative we apply this expression twice
\begin{eqnarray}
\frac{d^2  \calT_{\ee}}{d  \ee^2}  = \chi_{\ee}^{\star}
\left ( \left ( \pounds_{\vec{\Z}_{\ee}} + \frac{d}{d \ee}
\right ) \left ( \pounds_{\vec{\Z}_{\ee}} + \frac{d}{d \ee}
\right ) T_{\ee} \right )= \hspace{5cm} \nonumber \\
\chi_{\ee}^{\star} \left ( \pounds_{\vec{\Z}_{\ee}}
\pounds_{\vec{\Z}_{\ee}} T_{\ee} +
\pounds_{\vec{\Z}_{\ee}} \left ( \frac{d T_{\ee}}{d \ee}
\right ) +  
\frac{d \left (\pounds_{\vec{\Z}_{\ee}} T_{\ee}\right )} {d \ee} 
+ \frac{d^2 T_{\ee}}{d \ee^2} \right ).
\label{intermediate}
\end{eqnarray}
Only the third term needs to be elaborated.
Using again the definition
of derivative and adding and subtracting a suitable term
we get
\begin{eqnarray*}
\frac{d \left (\pounds_{\vec{\Z}_{\ee}} T_{\ee} \right )} {d
\ee} = \mathop{\lim} \limits_{h  \mathop   \to  0   }
\frac{\pounds_{\vec{\Z}_{\ee+h}} T_{\ee+h} -
\pounds_{\vec{\Z}_{\ee}} T_{\ee+h}}{h} + \mathop{\lim}
\limits_{h  \mathop   \to  0   }   \frac{\pounds_{\vec{\Z}_{\ee}}
T_{\ee+h} - \pounds_{\vec{\Z}_{\ee}} T_{\ee}}{h},
\end{eqnarray*}
which, using the linearity of the Lie derivative, becomes
\begin{eqnarray*}
\frac{d \left (\pounds_{\vec{\Z}_{\ee}} T_{\ee} \right )} {d
\ee} = \pounds_{\frac{d \vec{\Z}_{\ee}}{d \ee}}
T_{\ee} + \pounds_{\vec{\Z}_{\ee}} \left ( \frac{ d
T_{\ee}}{d \ee} \right )
\end{eqnarray*}
Inserting this into (\ref{intermediate}) and using the definition $
\vec{\W}_{\ee}  = \frac{d  \vec{\Z}_{\ee}}{d  \ee}$ 
the lemma follows. $\hfill \Box$

{\bf Remark.} Notice that when $\N$ is a hypersurface $\Sigma$,
all the information regarding the first and second variation
of the hypersurface $\Sigma_{\ee}$ around $\Sigma_{0}$ is encoded in the two vector fields 
$\vec{\Zone} \equiv \vec{Z}_{\ee=0}$ and $\vec{\W} \equiv \W_{\ee=0}$.
These vectors are, by construction, defined everywhere on $\M$
(because they are defined in terms of the diffeomorphisms $\Psi^{\ee}_h$).
However, only their values on $\Sigma_0$ should matter. Geometrically, they define 
how the hypersurface is deformed to first and second order (as we shall see, the second order variation
is best defined by using a suitable combination of $\vec{\W}$ and $\vec{\Zone}$).
These vectors have tangential and normal components. The normal part determines how the
hypersurface moves in spacetime as a set of points, while the tangential part encodes
the information on how the different $\Sigma_{\ee}$ are coordinated.
The fact that these vectors have been extended off $\Sigma_0$, and that this extension is 
essentially arbitrary (due to the large freedom in defining $\Psi_{h}^{\ee}$) provides
a powerful check for the validity of the final results, namely that they must be
independent of the extension  of these vectors outside $\Sigma_0$.

\section{First and second order perturbations vectors of $\Sigma$}
\label{Z1Z2}
Let us now concentrate in the case where $\N$ is a hypersurface $\Sigma$. We replace $\chi_{\ee} \rightarrow \Phi_{\ee}$
in all the expressions above. Our aim in this section is to find explicit expressions for the vectors 
$\Zone \equiv  \left .  \frac{\partial  \Psi_h^{\ee}}{
\partial  h}  \right  |_{\ee=h=0}$ and $\vec{\W}  \equiv \left . \frac{d  \vec{\Z}_{\ee}}{d  \ee} \right |_{\ee=0}$,
so that they can be determined in explicit examples. For the second variation we use 
$\vec{\Ztwo} \equiv \vec{\W} + \nabla_{\vec{\Zone}} \vec{\Zone}$ instead of $\vec{\W}$. The reason will
become clear later on.
We call $\vec{Z}_1$ and $\vec{\Ztwo}$
respectively as {\it first and second order perturbation vectors of $\Sigma$}. Let us choose
local coordinate systems on $\Sigma$ and on $\M$
so that $\Phi_{\ee}$ are written
\bestar
\Phi_{\ee}: \quad \Sigma & \longrightarrow & \M \\
            y^i & \longrightarrow & x^\alpha = \Phi^{\alpha}(y^i,\ee).
\enstar
\begin{proposition}
\label{1st2nd}
The first and second order perturbation vectors $\Zone^{\alpha}(y)$ and $\Ztwo^{\alpha}(y)$
of the hypersurface $\Sigma$ read
\begin{eqnarray}
\Zone^{\alpha} (y^i) \! & \! = \! & \!
\left . \frac{\partial \Phi^{\alpha} (y^i,\ee)}{\partial \ee} \right |_{\ee=0}, \label{Z1} \\
\Ztwo^{\alpha}(y^i) \! & \! = \! & \!
\left . \frac{\partial^2 \Phi^{\alpha} (y^i,\ee)}{\partial \ee^2 } \right |_{\ee=0}
+  \Gamma^{\alpha}_{\beta\gamma} (x_0(y^i)) \Zone^{\beta} (y^i) \Zone^{\gamma} (y^i).
\label{Z2}
\end{eqnarray}
where $x_0(y^i)$ is the local form of the unperturbed embedding $\Phi_0$.
\end{proposition}

{\it Proof:} $\Psi^{\ee}_h : \M \rightarrow \M$ satisfies $\Phi_{\ee+h} = \Psi^{\ee}_h \circ
\Phi_{\ee}$. Let $\Psi^{\alpha}(x^{\alpha},h,\ee)$ be the local coordinate form of $\Psi^{\ee}_h$.
By construction, $\Phi^{\alpha}$ and $\Psi^{\alpha}$ satisfy
\be 
\Phi^{\alpha}(y^i,\ee+h) = \Psi^{\alpha} (\Phi^{\beta}(y^i,\ee),h,\ee),
\label{RelationPhiPsi}
\en
which has as immediate consequence
\bestar
\left . \frac{\partial \Psi^{\alpha}}{\partial h} \right
|_{(\Phi(y^i,\ee),h,\ee)} = 
\left . \frac{\partial \Psi^{\alpha}}{\partial x^\beta} \right 
|_{(\Phi(y^i,\ee),h,\ee)} \left . 
\frac{\partial \Phi^{\beta}}{\partial \ee}  \right |_{(y^i,\ee)}
+ \left .  \frac{\partial \Psi^{\alpha}}{\partial \ee} \right |_{(\Phi(y^i,\ee),h,\ee)}.
\enstar
Evaluating at $\ee=h=0$ and using the fact that $\Psi^\ee_0$ is
the identity on $\M$ for all $\ee$ we obtain,
\be
\left . \frac{\partial \Psi^{\alpha}}{\partial h} \right |_{(x^{\alpha} =
x^{\alpha}_0(y^i),h=0,\ee=0)} = 
\left . \frac{\partial \Phi^{\alpha}(y^i,\ee)}{\partial \ee} \right |_{\ee=0}.
\label{Deffident}
\en
According to its definition, $\zee$ has components $Z^{\alpha}_{\ee}(x) = \left . \frac{\partial
\Psi^{\alpha}(x^{\beta},h,\ee)}{\partial h} \right |_{h=0}$. Expression (\ref{Z1})
follows directly from (\ref{Deffident}). For the second order perturbation vector
$\vec{\Ztwo}$, let us first find the coordinate
form of $\partial_{\ee} \zee + \nabla_{\zee} \zee$.  Directly from its definition 
one finds
\be
\left . \partial_{\ee} Z^{\alpha}_{\ee}(x) + \nabla_\zee Z_{\ee}^{\alpha}(x) \right |_{\ee=0} = 
\left . \frac{\partial^2 \Psi^{\alpha}}{ \partial h \partial \ee } \right |_{(x,0,0)} + 
\left . \frac{\partial \Psi^{\beta}}{ \partial h } \right |_{(x,0,0)} 
\left . \frac{\partial^2 \Psi^{\alpha}}{ \partial x^{\beta} \partial h } \right |_{(x,0,0)} +
\nonumber  \\
+ \Gamma^{\alpha}_{\beta\gamma} (x) \left . \frac{\partial \Psi^{\beta}}{ \partial h } 
\right |_{(x,0,0)} 
\left . \frac{\partial \Psi^{\gamma}}{ \partial h } 
\right |_{(x,0,0)}, \hspace{4cm}
\label{Z2t}
\en
which contains no second derivatives with respect to $h$. Performing 
the second $\ee,h$ derivative of (\ref{RelationPhiPsi}) and evaluating at $\ee=h=0$ gives
\be
\left . \frac{\partial^2 \Phi^{\alpha} (y^i,\ee+h)}{\partial \ee \partial h} \right |_{\ee=h=0} =
\left . \frac{\partial^2 \Psi^{\alpha}}{\partial x^{\beta} \partial h} \right |_{(x=x_0(y^i),0,0)}
\left . \frac{\partial \Phi^{\beta} (y^i,\ee)}{\partial \ee} \right |_{\ee=0} +
\left . \frac{\partial^2 \Psi^{\alpha}}{\partial h \partial \ee} \right |_{(x=x_0(y^i),0,0)}
\en
Taking into account 
$\left . \partial_{\ee} \partial_h \Phi^{\alpha} (y^i,\ee+h) \right |_{(\ee=0,h=0)} =
\left . \partial^2_{\ee} \Phi^{\alpha} (y^i,\ee+h) \right |_{(\ee=0,h=0)} $, and noticing that 
$\partial^2_\ee  \Phi^{\alpha} (y^i,\ee+h) |_{\ee=h=0} = 
\partial_\ee\partial_\ee \Phi^{\alpha} (y^i,\ee) |_{\ee=0}$, the vector field
$\partial_\ee Z^{\alpha}_\ee + \nabla_\zee Z_\ee^{\alpha}$  evaluated on $\Sigma$ and at $\ee=0$ becomes
\be
\left . \partial_\ee Z^{\alpha}_\ee + \nabla_\zee Z_\ee^{\alpha} \right |_{(x=x_0(y^i),\ee=0)} = 
\left . \frac{\partial^2 \Phi^{\alpha} (y^i,\ee)}{\partial \ee^2 } \right |_{\ee=0}
+  \Gamma^{\alpha}_{\beta\gamma} (x_0(y^i)) Z^{\beta} (y^i) Z^{\gamma} (y^i),
\label{Z2bis}
\en
and (\ref{Z2}) follows directly from its definition $\Ztwo^{\alpha} (y^i)  =
\left . \partial_\ee Z^{\alpha}_\ee + \nabla_\zee Z_\ee^{\alpha} \right |_{(x=x_0(y^i),\ee=0)}$
$\hfill \Box$

{\bf Remark:} Both the first and second order perturbation vectors $\vec{\Zone}$ and
$\vec{\Ztwo}$ depend only on the family of embeddings $\Phi_\ee$ (i.e. on the hypersurfaces
$\Sigma_{\ee}$) and not on the specific choice of 
$\Psi_h^{\ee}$.
Moreover, it  is clear from (\ref{Z2}) that 
$\Ztwo^{\alpha}(y^i)$ corresponds to the covariant acceleration,
evaluated on $\Sigma_0$, of the curve defined 
by the motion of a fixed point of $\Sigma$ when the hypersurface moves, i.e. by
the curve $\Phi^{\alpha}(y^i,\ee)$ with $y^i$ fixed. Notice also that, had we chosen
$\vec{W} = \partial_\ee \zee |_{\ee=0}$ as our second order perturbation vector, we would have obtained a
vector field which depends on  $\Psi_h^{\ee}$, i.e. it would not be defined solely in terms
of the one parameter family of embeddings. This is why $\vec{\Ztwo}$ is preferable to
$\vec{\W}$.

\section{First order perturbation of the hypersurface}

We can now start the calculations of the 
first and second order variations
of the fundamental forms of $\Sigma_0$.  We shall assume that 
this hypersurface contains no null points, i.e. 
that its induced first fundamental form $h = \Phi^{\star}_0 (g)$ 
defines a metric. 
For small enough $\epsilon$ the same will be true
for $\Sigma_{\ee}$ at least on compact subsets. Since we are only interested
in derivatives at $\ee=0$ we can assume without loss of generality that all 
 $\Sigma_{\ee}$ are non-degenerate. Let us denote by
$h_{\ee} = \Phi_{\ee}^{\star} (g)$ the one-parameter family of induced metrics. Notice that 
all of them are
defined on the same manifold $\Sigma$. Let also\footnote{In this paper boldface
letters are used to denote one-forms.} $\bm{n}_{\ee}$ 
be the unit normal to
$\Sigma_{\ee}$ with respect to $g_{\ee}$.
Its orientation is taken arbitrarily on $\Sigma_{0}$ an extended to all $\ee$
by continuity.  Let us extend $\bm{n}_{\ee}$
to an open neighbourhood ${\cal U}$ of $\Sigma_{\ee}$. By working locally near one point 
we can, without loss of generality, choose ${\cal U}$  to be independent of $\ee$.
We keep $\bm{n}_{\ee}$ unit everywhere on ${\cal U}$, i.e.
$g_{\ee \alpha \beta} n^{\alpha}_{\ee}
n^{\beta}_{\ee} |_{\U} = \sigma$, where $\sigma = +1$ for timelike hypersurfaces and $\sigma=-1$ for
spacelike ones. Hence, we have at hand a one-parameter family of one-forms $\bm{n}_{\ee}$ defined everywhere
on  $\U$
and both covariant derivatives at constant $\ee$ and $\ee$-derivatives at fixed spacetime point $x \in \U$ can
be performed. $\Sigma$ inherits a 
one-parameter family of second fundamental forms $\kappa_{\ee} = \Phi^{\star}
( \nabla^{\ee} \bm{n}_{\ee} )$. We drop the subindex $0$ for any background object, thus
$\Sigma_0$ is endowed with a metric $h$, 
covariant derivative $D$, second fundamental form $\kappa$ and has unit normal 
$\vec{n}$. We also write the background embedding simply as $\Phi$.
All spacetime indices are lowered and raised
with the background metric $g_{\alpha\beta}$ and it inverse. Similarly all hypersurface
indices are lowered and raised with $h_{ij}$ and
its inverse. Tensors on $\Sigma$ will carry Latin indices.

The first and second order perturbations of the induced metric and second fundamental forms are obviously
\begin{eqnarray*}
\mbox{First order perturbations:} & & \hspace{2cm} 
\qper = \left . \frac{\partial h_{\ee}}{\partial \ee}  \right |_{\ee=0}, \quad
\kappaper = \left . \frac{\partial \kappa_{\ee}}{\partial \ee}  \right |_{\ee=0}, \\
\mbox{Second order perturbations:} & & \hspace{2cm} 
\qperper = \left . \frac{\partial^2 h_{\ee}}{\partial \ee^2}  \right |_{\ee=0}, \quad
\kappaperper = \left . \frac{\partial^2 \kappa_{\ee}}{\partial \ee^2}  \right |_{\ee=0}, \\
\end{eqnarray*}
These derivatives are taken at fixed point $p$ in the abstract manifold $\Sigma$.
In this section we shall obtain the explicit expressions for the
first order perturbed quantities. The second order quantities are considered
in the next section.

Let us start by introducing some notation. It is well known that
covariant tensors on the background hypersurface $\Sigma$ are in one to one correspondence with
spacetime tensors defined on $\Sigma_0$ and which are totally tangent to $\Sigma_0$, i.e.
tensors $C_{\mun}$ satisfying $C_{\mun}n^{\mu_a} =0$ ($a=1 \cdots m$) on $\Sigma_0$. 
The one-form $\bm{n}$ is obviously hypersurface orthogonal on $\Sigma_0$. We
can assume without loss of generality that its extension off $\Sigma_0$ is chosen
so that this property is kept. We could also choose $\vec{n}$ so that it defines
a geodesic affinely parametrized congruence. These two conditions would fix $\vec{n}$ uniquely and
would simplify the calculations below. However we prefer to leave the acceleration
$\vec{a} = \nabla_{\vec{n}} \vec{n}$ completely free. The increase in complexity 
is compensated by the fact that the result {\it has to be independent of $\vec{a}$}. 
This provides a non-trivial check for the validity of the result.
Since, as we shall see, the calculation is quite involved,
it is convenient to keep non-trivial checks at hand. 

Being $\bm{n}$ hypersurface orthogonal, its 
covariant derivative reads
\begin{eqnarray}
\nabla_{\alpha} n_{\beta} = \sigma n_{\alpha} a_{\beta} + \kappa_{\alpha\beta}, \label{nabn}
\end{eqnarray}
where $\kappa_{\alpha\beta}$ is symmetric and completely orthogonal to $\vec{n}$.
This tensor is obviously the counterpart on $\M$ of the second fundamental form $\kappa_{ij}$.
From (\ref{nabn}) it follows
$\pounds_{\vec{n}} g_{\alpha\beta} = \sigma n_{\alpha} a_{\beta} + \sigma n_{\beta} a_{\alpha} + 
2 \kappa_{\alpha\beta}$.

Covariant derivatives of a tensor $C_{i_1 \cdots i_m}$
within the hypersurface can be calculated 
by considering its counterpart $C_{\mun}$ on spacetime. Indeed, if we extend this tensor to a neighbourhood of $\Sigma_0$ 
in such a way that it remains orthogonal to $\vec{n}$ (and otherwise arbitrarily), the three dimensional covariant
derivative can be calculated as
\begin{eqnarray*}
D_{\alpha} C_{\mun} \equiv h^{\nu}_{\alpha} h^{\beta_1}_{\mu_1} \cdots h^{\beta_m}_{\mu_m} 
\nabla_{\nu} C_{\beta_1 \cdots \beta_m},
\end{eqnarray*}
where $h^{\alpha}_{\beta} \equiv \delta^{\alpha}_{\beta} - \sigma n^{\alpha} n_{\beta}$ is the
projector to the hypersurface. More concretely, this means that
$D_j  C_{i_1 \cdots i_m}$ has $D_{\alpha}C_{\mun}$ as its spacetime counterpart.
A simple integration by parts shows that 
covariant derivatives and three-dimensional derivatives are related by
\begin{eqnarray}
\nabla_{\alpha} C_{\mun} = D_{\alpha} C_{\mun} + \sigma n_{\alpha} n^{\rho} \nabla_{\rho} C_{\mun}
- \sigma \sum_{i=1}^n C_{\murhon} \kappa^{\rho}_{\,\, \alpha} n_{\mu_i}.
\label{nablaD}
\end{eqnarray}

For later use let us notice some useful expressions. The first one
is obvious: for any covariant tensor $A_{\aln}$ and
any function $F$ 
\begin{eqnarray}
\left ( \pounds_{F \vec{n} } A \right )_{\aln} = F \left ( \pounds_{\vec{n}} A \right )_{\aln}
+ \sum_{i=1}^{n} A_{\almun} n^{\mu} \nabla_{\alpha_i} F.
\label{LieQn}
\end{eqnarray}
Less immediate, but still easy, are the following three Lemmas. The first
one is well-known
\begin{lemma}
\label{embed}
Let $\Sigma$ be an embedded  submanifold of $\M$ with embedding $\Phi : \Sigma \rightarrow \M$ 
and let $\vec{V}$ be a vector field on a neighbourhood $\U$ of $\Phi(\Sigma)$ tangent to this
hypersurface (i.e. 
$ \vec{V} |_{\Phi (\Sigma)} = \Phi_{\star} (\vec{V}_{\Sigma} )$ for some vector $\vec{V}_{\Sigma}$
on $\Sigma$).
Then, for any covariant tensor $A$ on $\U$
\begin{eqnarray*}
\Phi^{\star} \left ( \pounds_{\vec{V}} A \right ) = \pounds_{\vec{V}_{\Sigma}} \left ( \Phi^{\star} A \right ).
\end{eqnarray*}
\end{lemma}
{\bf Remark.} For simplicity  we will use the same symbol to denote $\vec{V}_{\Sigma}$ and $\vec{V}$.
The precise meaning will be clear from the context.

A consequence of this Lemma is that, for any vector field $\vec{\X} = R \vec{n} + \vec{V}$, with
$\vec{V}$ orthogonal to $\vec{n}$, 
\begin{eqnarray}
\Phi^{\star} \left ( \pounds_{\vec{\X}} g \right ) =
\pounds_{\vec{V}} \, h + 2 R \kappa,
\label{LieRn}
\end{eqnarray}

\begin{lemma}
\label{LieS}
Let $B_{\alpha\beta}$ be any symmetric tensor and 
$\vec{\X}$ any vector field. 
Defining $S(B)^{\mu}_{\alpha\beta} \equiv \frac{1}{2}
(\nabla_{\alpha} B^{\mu}_{\,\, \beta} + \nabla_{\beta} B^{\mu}_{\,\,
\alpha} - \nabla^{\mu} B_{\alpha\beta})$
and $H^{\alpha} = B^{\alpha\mu} \X_{\mu}$. The following identity holds
\begin{eqnarray*}
\left ( \pounds_{\vec{\X}} B \right )_{\alpha\beta}  + 2 \X_{\mu} S(B)^{\mu}_{\alpha\beta} =
\left ( \pounds_{\vec{H}} g \right )_{\alpha\beta}.
\end{eqnarray*}
\end{lemma}
The next Lemma tells us how to perform second Lie derivatives.
The Riemann tensor of $(\M,g)$ is denoted by $R_{\alpha\beta\gamma\delta}$.
\begin{lemma}
\label{liewwcov} 
Let $\vec{\X}$ be an arbitrary vector field and $B_{\alpha\beta}$ 
any symmetric tensor.
Then
\begin{eqnarray*}
\left (\pounds_{\vec{\X}} \pounds_{\vec{\X}} B \right )_{\alpha \beta} = 
\left ( \pounds_{\nabla_{\vec{\X}} \vec{\X}} B
\right )_{\alpha\beta} + \X^{\mu} \X^{\nu} \nabla_{\mu} \nabla_{\nu} B_{\alpha\beta} - 
B_{\alpha \nu} \X^{\mu} \X^{\gamma} R^{\nu}_{\,\,\, \gamma \beta \mu} -
B_{\beta \nu} \X^{\mu} \X^{\gamma} R^{\nu}_{\,\,\, \gamma \alpha \mu}  \nonumber  \\ 
+ 2 \left ( \X^{\mu} \nabla_{\mu} B_{\alpha \nu}  \right) \nabla_{\beta} \X^{\nu} + 
2 \left ( \X^{\mu} \nabla_{\mu} B_{\beta \nu}  \right) \nabla_{\alpha} \X^{\nu} + 
2 B_{\mu\nu} \left ( \nabla_{\alpha} \X^{\mu} \right ) \left ( \nabla_{\beta} \X^{\nu} \right ),
\end{eqnarray*}
\end{lemma}
{\it Proof:} Expand the first term and use the Ricci identity applied
to $\vec{\X}$. $\hfill \Box$

A particular case of this Lemma is obtained for $B_{\alpha\beta} = g_{\alpha\beta}$:
\begin{eqnarray}
\left (\pounds_{\vec{\X}} \pounds_{\vec{\X}} g \right )_{\alpha \beta} = ( \pounds_{\nabla_{\vec{\X}} \vec{\X}} g
)_{\alpha\beta} - 2 \X^{\mu} \X^{\nu} R_{\alpha \mu \beta \nu}
+ 2 \left ( \nabla_{\alpha} \X^{\mu} \right ) \left ( \nabla_{\beta} \X_{\mu} \right ).
\label{LieXXg}
\end{eqnarray}
Combining this with the general expression (\ref{LieQn}) it follows easily that,  
for any pair of functions $F_1$ and $F_2$,
\begin{eqnarray}
\pounds_{F_1 \vec{n}} \pounds_{F_2 \vec{n}} \, g_{\alpha\beta} =
\pounds_{F_1 F_2 \vec{a}} \, g_{\alpha\beta} 
+ 2 F_1 F_2 \left ( -  n^{\mu} n^{\nu}
R_{\alpha\mu\beta\nu} +  \kappa_{\alpha\mu} \kappa_{\beta}^{\,\,\,\mu} \right ) \nonumber \\
+ \sigma \left ( \nabla_{\alpha} F_1 \nabla_{\beta} F_2 +
\nabla_{\alpha} F_2 \nabla_{\beta} F_1 \right ) 
+ 2 \kappa_{\alpha\beta} F_1 \vec{n} (F_2 ) + n_{\alpha}
G_{\beta} + n_{\beta} G_{\alpha}, \label{LieF1F2g}
\end{eqnarray}  
where $G_{\alpha} = \sigma a_{\alpha} F_1 \vec{n} ( F_2 )
+ \pounds_{F_1 \vec{n}} (\nabla_{\alpha} F_2)+
2 \sigma F_1 F_2  a^{\mu} \kappa_{\mu\alpha}
+ F_1 F_2 n_{\alpha} a^{\mu} a_{\mu}$. With $F_1 \rightarrow \Qone$
and $F_2 \rightarrow 1$ we get 
\begin{eqnarray}
\pounds_{\Qone\vec{n}} \pounds_{\vec{n}} g_{\alpha\beta} & =& 
\pounds_{\Qone \vec{a}} g_{\alpha\beta}+ \nonumber \\
& + &  2 \Qone \left (
-  n^{\mu} n^{\nu}R_{\alpha\mu\beta\nu} 
+ \kappa_{\alpha\mu} \kappa_{\beta}^{\,\,\,\mu} 
+ n_{\alpha} n_{\beta} a_{\mu} a^{\mu} 
+ \sigma n_{\alpha} a^{\mu} \kappa_{\beta\mu} 
+ \sigma n_{\beta} a^{\mu} \kappa_{\alpha\mu} \right )
\label{LieQ1g}
\end{eqnarray}  
It is convenient to decompose
the first perturbed metric into tangential and normal components with respect to $\vec{n}$. Explicitly
\begin{eqnarray}
\Kper_{\alpha \beta} = \Kpernornor n_{\alpha} n_{\beta} + \sigma n_{\alpha} \Kpernortan_{\beta} +
\sigma n_{\beta} \Kpernortan_{\alpha} + \Kpertan_{\alpha\beta},
\label{decomKper}
\end{eqnarray}
where obviously $\Kpernortan_{\alpha}$ and $\Kpertan_{\alpha\beta}$ are
orthogonal to $n^{\alpha}$.
We can now find the first order perturbations of the first and second
fundamental forms of $\Sigma_0$. 
\begin{proposition}[Battye \& Carter, 1995]
\label{FirstPert}
Let $(\M,g)$ be a $C^3$ spacetime of any dimension and $\Sigma_0$ an arbitrary non-degenerate hypersurface 
defined by an embedding $\Phi: \Sigma \rightarrow \M$. Let $h$ be the induced
metric, $\kappa$ the extrinsic curvature and $\vec{n}$ the unit normal to the hypersurface.

If the metric $g$ is perturbed to first order with
$\Kper$ and the hypersurface is perturbed to first order
with a vector field $\vec{\Zone} = \Qone \vec{n} + \vec{\Tone}$, where $\vec{\Tone}$ is tangent to 
$\Sigma_0$, then
the induced metric and extrinsic curvature are perturbed to first order as
\begin{eqnarray}
\qper_{ij} & \! \! = \! \! &  
\pounds_{\vec{\Tone} } h_{ij}  + 2 \Qone \kappa_{ij} + \Kper_{\alpha\beta} e^{\alpha}_i e^{\beta}_j,
\label{FirstPert.1}
\\
\kappaper_{ij} & \! \! = \! \! &  
\pounds_{\vec{\Tone} } \kappa_{ij} 
- \sigma D_{i} D_{j} \Qone 
+ \Qone \left ( - n^{\mu} n^{\nu} R_{\alpha\mu\beta\nu}
e^{\alpha}_i e^{\beta}_j + \kappa_{il} \kappa^{\,\,l}_{j} \right )
+ \frac{\sigma}{2}  \Kpernornor 
\kappa_{ij} - n_{\mu} \Sper^{\mu}_{\alpha\beta}e^{\alpha}_i e^{\beta}_j, \nonumber
\end{eqnarray}
where $\Kpernornor = \Kper_{\alpha\beta} n^{\alpha} n^{\beta}$, $\Sper$ is given
in (\ref{Sper}) and $e^{\alpha}_i = \Phi_{\star} (\partial_{i})$ are tangent vectors to $\Sigma_0$.
\end{proposition}
{\it Proof:} From Lemma \ref{FirstSecondOrder} 
\begin{eqnarray}
\qper = \partial_{\ee} h_{\ee} |_{\ee=0} = 
\Phi^{\star} \left ( \pounds_{\vec{\Zone} } g \right ) + \Phi^{\star} \left ( \Kper \right )  
\label{Lemma1qper}
\end{eqnarray}
and  (\ref{FirstPert.1})
follows directly from Lemma \ref{embed} and (\ref{LieRn}). 
For $\kappaper$, we notice that $2 \kappa_{\ee} = 2 \Phi^{\star}_{\ee} ( \nabla^{\ee} \bm{n_{\ee}} )
= \Phi^{\star}_{\ee} ( \pounds_{\vec{n}_{\ee}} g_{\ee} )$. Applying Lemma \ref{FirstSecondOrder}
\begin{eqnarray}
\left . 2 \partial_{\ee} \kappa_{\ee}  \right |_{\ee=0} = 
\Phi^{\star} \left ( \pounds_{\vec{\Zone}} \pounds_{\vec{n}} g + \pounds_{\vec{\none}}
g + \pounds_{\vec{n}} \Kper 
\right ), 
\label{kappaper.1}
\end{eqnarray}
where $\vec{\none} \equiv  \left . \partial_{\ee} \vec{n}_{\ee} \right |_{\ee = 0}$. Let us
identify this vector: for its  normal component we use that 
$\vec{n}_{\ee}$  is unit for all $\ee$, i.e.
$( \vec{n}_{\ee} , \vec{n}_{\ee} )_{\gee} = \sigma$. The derivative 
at $\ee=0$ gives 
$\none^{\alpha} n_{\alpha} =- \frac{1}{2} \Kpernornor$.
For its tangential part, it is convenient to use the normal one-form,
$\bm{\mone} \equiv \partial_{\ee} \bm{n}_{\ee} |_{\ee=0}$. 
From  $\Phi^{\star}_{\ee} ( \bm{n_{\ee}} ) =0$, Lemma \ref{FirstSecondOrder} gives
$\Phi^{\star} ( \pounds_{\vec{\Zone}} \, \bm{n} + \bm{\mone}  ) = 0$.
Using $\pounds_{\vec{n}} ( \bm{n} )_{\alpha}  =  a_{\alpha}$ and
(\ref{LieQn}) with $A = \bm{n}$, $F = \Qone$ gives
\begin{eqnarray}
\Phi^{\star} ( \bm{\mone} )_{i} = - \left ( \Qone a_i + \sigma D_i \Qone \right ). 
\label{tangentialmone}
\end{eqnarray}
The identity $\partial_{\ee} g^{\alpha\beta}_{\ee} = - g^{\alpha\mu}_{\ee}
g_{\ee}^{\beta \nu} \partial_{\ee} g_{\ee \,\mu\nu}$ implies that 
$\vec{\none}$ and $\vec{\mone}$ are related by
$\none^{\alpha} = - \Kper^{\alpha \beta} n_{\beta} + \mone^{\alpha}$. Hence
$\mone_{\alpha} n^{\alpha} = \frac{1}{2} \Kpernornor$ which, together with (\ref{tangentialmone}),
gives
\begin{eqnarray}
\mone^{\alpha} =  \frac{\sigma}{2}  \Kpernornor n^{\alpha} -
\left ( \Qone a^{\alpha} + \sigma D^{\alpha} \Qone \right ),
\label{mone} \\
\none^{\alpha} = - \frac{\sigma}{2}  \Kpernornor n^{\alpha} - \left (
\Kpernortan^{\alpha} +
\Qone a^{\alpha} + \sigma D^{\alpha} \Qone \right ),
\label{none} 
\end{eqnarray}
where the decomposition (\ref{decomKper}) has been used.
Inserting (\ref{none}) and (\ref{LieQ1g}) into (\ref{kappaper.1}) yields 
\begin{eqnarray}
\left . 2 \partial_{\ee} \kappa_{\ee \, ij}  \right |_{\ee=0} = 2 \pounds_{\vec{\Tone} } \kappa_{ij} 
+ \pounds_{\Qone \vec{a}} \, h_{ij} 
+ 2 \Qone \left ( - n^{\mu} n^{\nu} R_{\alpha\mu\beta\nu}
e^{\alpha}_i e^{\beta}_j + \kappa_{il} \kappa^{\,\,l}_{j} \right ) - \sigma \Kpernornor \kappa_{ij} 
\nonumber \\
- \pounds_{ \vec{\Kpernortan} + \Qone \vec{a} + \sigma
  \grad \Qone } \, h_{ij}
+ \Phi^{\star} \left ( \pounds_{\vec{n}} \Kper \right ),
\label{kappaper.2}
\end{eqnarray}
where $(\gradL  \Qone) ^i = D^i \Qone$. For the  last term we apply Lemma
\ref{LieS} with $B_{\alpha\beta} = \Kper_{\alpha\beta}$
and $\vec{\X} = \vec{n}$, i.e.
\begin{eqnarray}
\left ( \pounds_{\vec{n}} \Kper \right )_{\alpha\beta} = 
- 2 n_{\mu} \Sper^{\mu}_{\alpha\beta} + \pounds_{\vec{\Kpernortan}
+ \sigma \Kpernornor \vec{n} } \, g_{\alpha\beta}
\label{LieKperSper}
\end{eqnarray}
and the Lemma follows. $\hfill \Box$

{\bf Remark.} This result was also derived by Mukohyama \cite{Mukohyama00} in his study of first order
perturbed matching conditions between spacetimes. The proof presented here is based on Lemma \ref{FirstSecondOrder}
which makes the calculations very efficient. This will allow
us to push the calculation to second order.

\section{Second order perturbation of the hypersurface}

Lemma \ref{liewwcov} gives an expression for second order
Lie derivatives. The next Lemma gives an alternative expression adapted to decompositions into tangential 
and normal components to the hypersurface.
\begin{lemma}
\label{LieZZ}
Let $\Sigma_0$ be an arbitrary non-generate hypersurface in a spacetime $(\M,g)$. Let $\vec{n}$
be a hypersurface orthogonal unit vector in a neighbourhood of $\Sigma_0$ which is orthogonal to $\Sigma_0$. Let $B$ be any
tensor  and $\vec{\Zone} = \Qone \vec{n} + \vec{\Tone}$, with $\vec{\Tone}$ orthogonal to
$\vec{n}$. Then
\begin{eqnarray}
\label{LieZZBbis}
\pounds_{\vec{\Zone}} \pounds_{\vec{\Zone}} B = 
\pounds_{\nabla_{\vec{\Zone}} \vec{\Zone}} B +
\pounds_{\vec{\Tone}} \pounds_{\vec{\Tone}} B + 2 \pounds_{\vec{\Tone}} 
\pounds_{\Qone \vec{n}} B - \pounds_{\Cone \vec{n} + \vec{\Done}} B - \pounds_{\Qone^2 \vec{a}} B
+ \pounds_{\Qone\vec{n}} \pounds_{\Qone\vec{n}} B,
\end{eqnarray}
where 
\begin{eqnarray}
\Cone
\equiv \Qone \vec{n} (\Qone)  + 2 \vec{\Tone} (\Qone) - \sigma \Tone^{\alpha} \Tone^{\beta} \kappa_{\alpha\beta}, \quad
\Done^{\mu} = 2 \Qone \Tone^{\alpha} \kappa_{\alpha}^{\,\, \mu}  + \Tone^{\alpha} D_{\alpha} \Tone^{\mu}.
\label{Cone}
\end{eqnarray}
\end{lemma}
{\it Proof:} 
Applying (\ref{nablaD}) to $\Tone_{\mu}$ in the decomposition $\vec{\Zone} = \Qone \vec{n} + \vec{\Tone}$, we get
\begin{eqnarray}
\nabla_{\alpha} \Zone_{\mu} = \sigma n_{\alpha} \left ( \Qone a_{\mu} + n^{\rho} \nabla_{\rho} \Tone_{\mu} \right )
+ n_{\mu} \left ( \nabla_{\alpha} \Qone - \sigma \Tone^{\rho} \kappa_{\rho \alpha} \right )
+ \Qone \kappa_{\alpha \mu} + D_{\alpha} \Tone_{\mu}. \label{nabZ}
\end{eqnarray}
Furthermore, linearity and the general property $\pounds_{\vec{X}} \pounds_{\vec{Y}} -\pounds_{\vec{Y}}
\pounds_{\vec{X}} = \pounds_{[\vec{X}, \vec{Y}]}$ imply
\begin{eqnarray}
\pounds_{\vec{\Zone}} \pounds_{\vec{\Zone}} B = \pounds_{\vec{\Tone}} \pounds_{\vec{\Tone}} B + 2 \pounds_{\vec{\Tone}} 
\pounds_{\Qone \vec{n}} B + \pounds_{[\Qone \vec{n}, \vec{\Tone}]} B + \pounds_{\Qone\vec{n}} \pounds_{\Qone\vec{n}} B.
\label{LieZZB}
\end{eqnarray}
We want to introduce
a term $\nabla_{\vec{\Zone}} \vec{\Zone}$ in the right hand side. To that aim
we rewrite $[\Qone \vec{n}, \vec{\Tone}]$ in the following way  
\begin{eqnarray}
[ \Qone \vec{n}, \vec{\Tone}]^{\mu} = 
[ \vec{\Zone}, \vec{\Tone}]^{\mu} =
\nabla_{\vec{\Zone}} \Tone^{\mu} - 
\nabla_{\vec{\Tone}} \Zone^{\mu} =
\nabla_{\vec{\Zone}} \Zone^{\mu} - 
\nabla_{\vec{\Zone}} \left ( \Qone n^{\mu} \right ) -
\nabla_{\vec{\Tone}}  \Zone^{\mu} = \nonumber \\  
\nabla_{\vec{\Zone}} \Zone^{\mu} - \Qone^2 a^{\mu} 
- 2 \Qone \Tone^{\alpha} \kappa_{\alpha}^{\,\, \mu}  - \Tone^{\alpha} D_{\alpha} \Tone^{\mu} 
+ n^{\mu} \left ( - \Qone \vec{n} (\Qone) - 2 \vec{\Tone} (\Qone) +
\sigma \Tone^{\alpha} 
\Tone^{\beta} \kappa_{\alpha\beta} \right ),
\label{QnT}
\end{eqnarray}
where the first three equalities are immediate and the last one 
follows directly from (\ref{nabn}) and (\ref{nabZ}).
Combining (\ref{LieZZB}) and (\ref{QnT}) the Lemma  follows.
$\hfill \Box$

As  before, let us decompose the second
order perturbed metric into tangential and normal components 
\begin{eqnarray}
\Kperper_{\alpha \beta} = \Kperpernornor n_{\alpha} n_{\beta} + \sigma n_{\alpha} \Kperpernortan_{\beta} +
\sigma n_{\beta} \Kperpernortan_{\alpha} + \Kperpertan_{\alpha\beta}.
\label{decomKperper}
\end{eqnarray}
We can now write down our main result of this section. The proof is given in Appendix A.
\begin{proposition}
\label{SecondPert}
With the same assumptions and notation as in Proposition \ref{FirstPert}, if the metric
is perturbed to second order with $\Kperper$ and the hypersurface is 
perturbed to second order with $\vec{\Ztwo} = \Qtwo \vec{n} + \vec{\Ttwo}$ (with $\vec{\Ttwo}$ orthogonal to 
$\vec{n}$) then the induced metric and
extrinsic curvature are perturbed to second order as
\begin{eqnarray}
\qperper_{ij} = \pounds_{\vec{\Ttwo}} h_{ij} + 2 \Qtwo \kappa_{ij} + \Kperper_{\alpha\beta} e^{\alpha}_i e^{\beta}_j
+ 2 \pounds_{\vec{\Tone}} \qper_{ij} - \pounds_{\vec{\Tone}} \pounds_{\vec{\Tone}} h_{ij} + \nonumber \\
+ \pounds_{2 \Qone \vec{\Kpernortan} - 2 \Qone \kappa (\vec{\Tone} )   -  D_{\vec{\Tone}} \vec{\Tone} } h_{ij} 
+ 2 \left (\sigma \Tone^{l}  \Tone^{s} \kappa_{ls}
- 2 \vec{\Tone} ( \Qone )  + 2 \sigma \Qone \Kpernornor \right )
\kappa_{ij} + \nonumber \\
+ 2 \Qone^2 \left ( -n^{\mu} n^{\nu} R_{\alpha\mu\beta\nu} e^{\alpha}_i
e^{\beta}_j + \kappa_{il} \kappa^{l}_{j} \right ) + 2 \sigma D_{i} \Qone D_{j} \Qone
- 4 \Qone n_{\mu} \Sper^{\mu}_{\alpha\beta} e^{\alpha}_i e^{\beta}_j
, \label{qperper}\\
\nonumber \\
\kappaperper_{ij} =
\pounds_{\vec{\Ttwo}} \kappa_{ij} - \sigma D_{i} D_{j} \Qtwo 
- \Qtwo n^{\mu} n^{\nu} R_{\alpha\mu\beta\nu} e^{\alpha}_i e^{\beta}_j 
+ \Qtwo \kappa_{il} \kappa^{l}_{k} 
- n_{\mu} \Sperper^{\mu}_{\alpha\beta} e^{\alpha}_i e^{\beta}_j 
 \nonumber \\
+ 2 \pounds_{\vec{\Tone}} \kappaper_{ij}
+ \kappa_{ij} \left ( \frac{\sigma}{2} \Kperpernornor - \frac{1}{4} \Kpernornor^2 - \sigma 
\left ( \tau_{l} + \sigma D_{l}\Qone \right )
\left ( \tau^{l} + \sigma D^{l}\Qone \right ) 
+ 2 \sigma \Qone n_{\mu} n^{\rho} n^{\delta}\Sper^{\mu}_{\rho\delta} \right )
\nonumber \\
+  \left ( \sigma \Kpernornor n_{\mu} + 2 \tau_{\mu} + 2 \sigma D_{\mu} \Qone \right ) 
\Sper^{\mu}_{\alpha\beta} e^{\alpha}_i e^{\beta}_j
- 2 \Qone n_{\mu} n^{\nu} ( \nabla_{\nu} \Sper^{\mu}_{\alpha\beta} ) e^{\alpha}_i e^{\beta}_j
- 2 n_{\mu} n^{\nu} \Sper^{\mu}_{\alpha \nu} e^{\alpha}_i D_j \Qone \nonumber \\
- 2 n_{\mu} n^{\nu} \Sper^{\mu}_{\alpha \nu} e^{\alpha}_j D_i \Qone 
- 2 \Qone n_{\mu} S^{\mu}_{\alpha \beta} e^{\alpha}_i e^{\beta}_l \kappa^{l}_j
- 2 \Qone n_{\mu} S^{\mu}_{\alpha \beta} e^{\alpha}_j e^{\beta}_l \kappa^{l}_i
\nonumber \\ 
+ \pounds_{\sigma \grad (\vec{\Tone} (\Qone)) 
- \frac{1}{2} \grad (\Tone^{l} \Tone^{m} \kappa_{lm} )
- \frac{1}{2} \Kpernornor \grad (\Qone) + 2 \sigma \Qone \kappa ( \grad \Qone )} \,  h_{ij}
\nonumber \\ 
+  \left ( 2 \vec{\Tone} (\Qone) - \sigma \Tone^{l} \Tone^{m} \kappa_{lm} -
\sigma \Qone \Kpernornor \right )\left ( n^{\mu} n^{\nu} R_{\alpha\mu\beta\nu}
e^{\alpha}_i e^{\beta}_j - \kappa_{il} \kappa^{l}_{j}  \right ) 
+ \frac{1}{2} \left ( D_{i} \Qone D_j \Kpernornor + \right . \nonumber \\
\left. + D_{j} \Qone D_i \Kpernornor \right ) 
-  \pounds_{\vec{\Tone}} \pounds_{\vec{\Tone}} \kappa_{ij} 
-  \pounds_{2 \Qone \kappa (\vec{\Tone} ) +  D_{\vec{\Tone}} \vec{\Tone} } \, \kappa_{ij} 
- 2 \sigma \Qone \pounds_{\grad (\Qone) } \kappa_{ij}
\nonumber \\
- \Qone^2 \left ( 
n^{\mu} n^{\nu} n^{\delta} ( \nabla_{\delta} R_{\alpha\mu\beta\nu} ) e^{\alpha}_i e^{\beta}_j
+ 2 n^{\mu} n^{\nu} R_{\delta\mu\alpha\nu} e^{\delta}_l e^{\alpha}_j \kappa^{l}_i
+ 2 n^{\mu} n^{\nu} R_{\delta\mu\alpha\nu} e^{\delta}_l e^{\alpha}_i \kappa^{l}_j \right ),
\label{kappaperper}
\end{eqnarray}
where $\Sperper$ is given in (\ref{Sperper}) and, for any tangent vector $\vec{V}$,
$\kappa(\vec{V})$ is the vector with components $\kappa^{i}_{\, j} V^j$,

\end{proposition}

{\bf Remark:} Propositions \ref{FirstPert}  and \ref{SecondPert} are still true for metrics $g$
of arbitrary signature.

The expressions in this Proposition are rather involved, and the calculations
leading to them are not easy. Thus, it is useful to have ways of checking whether
the expressions are indeed correct. We have already mentioned two such tests, namely that the result
must be completely independent of the extension of the normal $\bm{n}$ and of the perturbations vectors
$\vec{\Zone}$ and $\vec{\Ztwo}$ outside $\Sigma_0$. This is obvious from the expressions above because no term
containing the acceleration $\vec{a}$ is present. Moreover, there are no normal derivatives of objects defined
intrinsically on $\Sigma_0$. Another
important test is that the expressions must transform correctly under gauge transformations.
In the next section we study gauge transformations when there is a hypersurface present.

\section{First and second order gauge transformations}

The definition of the perturbation of the metric and of the hypersurface
is based on an a priori identification of $\M_{\ee}$ with $\M_0 = \M$ 
and of $\Sigma_{\ee}$ with the abstract manifold $\Sigma$. Obviously this identification is not
unique. The consequence of this non-uniqueness if the gauge freedom  inherent to any 
geometric perturbation theory. In our case we have two different sources of gauge
freedom, namely the identification of the ambient space and the identification of the hypersurfaces.
Let us describe them in detail.

The freedom in the identification of $\M_{\ee}$ is described by the possibility
of performing an arbitrary diffeomorphism of $\M_{\ee}$ (which, of course, will 
in general depend on $\ee$) before doing the identification with $\M = \M_{0}$.
An equivalent way of stating this fact is that for any one-parameter family of metrics $g_{\ee}$ on $\M$
we can define an infinite number of equivalent families by  performing an $\ee$-diffeomorphism
of $g_{\ee}$. The equivalence is shown as follows. Each  $\M_{\ee}$ has a
metric that we denote by $\hat{g}_{\ee}$. These metrics cannot be compared with each other
because they are defined on different spaces. Thus, we need a one-parameter family of diffeomorphisms
$\AA_{\ee} : \M \rightarrow \M_{\ee}$ in order to be able to relate them, and this defines
a one-parameter family of metrics on $\M$ by $g_{\ee} = \AA_{\ee}^{\star} (\hat{g}_{\ee})$.
If we now perform a diffeomorphism
$\Omega_{\ee}$ on $\M$ before identifying with $\M_{\ee}$, it is clear that $\AA_{\ee}$ becomes
$\AA^{(g)}_{\ee} = \AA_{\ee} \circ \Omega_{\ee}$ (the superscript $(g)$ stands for ``gauge transformed''),
and the new parameter family of metrics is $g^{(g)}_{\ee} = {\AA^{(g)}}^{\star} (\hat{g}_{\ee} ) =
\Omega^{\star}_{\ee} (g_{\ee})$, as claimed. All these families are geometrically equivalent. So, although
the  first  and  second  order  perturbation  metrics
obtained from  $g^{(g)}_{\ee}$ and $g_{\ee}$ are indeed different, 
they are intrinsically the same
(i.e. gauge equivalent). In this section we use 
Lemma  \ref{FirstSecondOrder} 
to find the gauge transformation
law for  $\Kper_{\alpha\beta}$  and $\Kperper_{\alpha\beta}$ 
(c.f.  \cite{Bruni97} and \cite{Sonego98} where the so-called knight diffeomorphism are
used to describe gauge transformations of any order).

Let  us  then apply  Lemma  \ref{FirstSecondOrder}  with $\N  =  \M$  and
$\chi_{\ee}    =    \Omega_{\ee}$.  In this
case  the    diffeomorphisms
$\Psi_{h}^{\ee}$  which   make  the  diagram  (\ref{DiagramComm})
commutative are  uniquely given by $\Psi_{h}^{\ee}  = \Omega_{h +
\ee} \circ \Omega^{-1}_{\ee}$. Choose $T_{\ee}
=  g_{\ee}$  so that  $\calT_{\ee} =  g^{(g)}_{\ee}$. We
write,   as   usual,    $\Kper   =   \frac{d   g_{\ee}}{d   \ee}
|_{\ee=0},    \Kperper   =   \frac{d^2    g_{\ee}}{d   \ee^2}
|_{\ee=0}$ and we define their gauge transformed tensors
as 
\begin{eqnarray*}
\Kpergauge  \equiv \left  . \frac{d  g^{(g)}_{\ee}}{d  \ee} \right
|_{\ee=0},
\hspace{2cm} \Kperpergauge \equiv  \left . \frac{d^2 g^{(g)}_{\ee}}{d \ee^2}
\right |_{\ee=0}.
\end{eqnarray*}
Let us also define
\begin{eqnarray}
\vec{\Sone} = \left.
\frac{\partial \Omega_{\ee}}{\partial \ee}  \right |_{\ee=0}, \hspace{5mm}
\vec{S}_{\ee} = \left . \frac{\partial \left ( \Omega_{h+\ee} \circ \Omega^{-1}_{\ee} 
\right )}{\partial h} \right |_{h=0}, \hspace{5mm}
\vec{V} = \left .\frac{\partial \vec{S}_{\ee}}{\partial \ee} \right |_{\ee=0}
\hspace{5mm}
\vec{\Stwo} = \vec{V} + \nabla_{\vec{\Sone}} \vec{\Sone} \label{GaugeVectors}
\end{eqnarray}
and call $\vec{\Sone}$ and $\vec{\Stwo}$ the {\it first and second order gauge vectors}.
\begin{proposition}[Bruni {\it \bf et. al.}, 1997]
\label{Gauge}
Under a gauge transformation defined by the vectors $\vec{\Sone}$ and $\vec{\Stwo}$, the first and
second order perturbed metric transform as
\begin{eqnarray*}
\Kpergauge_{\alpha\beta} & = & \Kper_{\alpha\beta} + \pounds_{\vec{\Sone}} g_{\alpha\beta}, \\
\Kperpergauge_{\alpha\beta} & = & \Kperper_{\alpha\beta} + \pounds_{\vec{\Stwo}} g_{\alpha\beta} +
2 \pounds_{\vec{\Sone}} \Kper_{\alpha\beta}
- 2 \Sone^{\mu} \Sone^{\nu} R_{\alpha\mu\beta\nu}
+ 2 \nabla_{\alpha} \Sone^{\mu}\nabla_{\beta} \Sone_{\mu}.
\end{eqnarray*}
\end{proposition}
{\it Proof:} The first expression is a direct consequence of Lemma \ref{FirstSecondOrder}. 
For the second order perturbation, the same Lemma gives
\begin{eqnarray*}
\Kperpergauge= \pounds_{\vec{V}} g+\pounds_{\vec{\Sone}}
\pounds_{\vec{\Sone}} g + 2
\pounds_{\vec{\Sone}} \Kper + \Kperper.
\end{eqnarray*}
The Proposition follows directly from  Lemma \ref{liewwcov} with $\vec{X} \rightarrow \vec{\Sone}$. $\hfill \Box$

{\bf Remark:} In \cite{Bruni97} the vector $\vec{V}$ was used instead of $\vec{\Stwo}$ to define 
second order gauge transformations. Both vectors are equally suited in this
case (unlike for hypersurfaces, where  $\vec{\Ztwo}$
is intrinsic to the perturbation $\Sigma$ while $\vec{\W}$ is not, see Remark after Proposition
\ref{1st2nd}). Second order gauge transformation
were analyzed also in \cite{Battye00} and
$\vec{\Stwo}$ was used there.

When performing a gauge transformation not only the perturbed metrics change but also the
vectors $\vec{\Zone}$ and $\vec{\Ztwo}$ of the hypersurface are modified. Geometrically, this
is clear because changing the way how the different manifolds $\M_{\ee}$ are
identified to each other affects how the abstract manifold $\Sigma$ is embedded into $\M$ 
at different $\ee$, and this obviously changes $\vec{\Zone}$ and $\vec{\Ztwo}$. It is clear, for instance, that
in suitably chosen gauges one can always make these two vectors identically zero.
Let us therefore determine the behaviour
of $\vec{\Zone}$ and $\vec{\Ztwo}$ under general gauge transformations.
We denote $\vec{\Zone}^{(g)}$ and $\vec{\Ztwo}^{(g)}$ the corresponding gauge transformed
vectors. 
\begin{proposition}
\label{ZGauge}
Under a gauge transformation defined by $\vec{\Sone}$
and $\vec{\Stwo}$, the first and second
order perturbation vectors of $\Sigma$ transform as
\begin{eqnarray}
\vec{\Zone}^{(g)} = \vec{\Zone} - \vec{\Sone}, \nonumber \\
\vec{\Ztwo}^{(g)} = \vec{\Ztwo} - \vec{\Stwo} - 2 \nabla_{\vec{\Zone}} \vec{\Sone} + 2 \nabla_{\vec{\Sone}} 
\vec{\Sone}. \label{gauge2}
\end{eqnarray}
\end{proposition}
{\it Proof:} The hypersurface $\Sigma_{\ee}$ is intrinsically (i.e. gauge independently) embedded
into $\M_{\ee}$. However, we are viewing them as hypersurfaces of $\M$ via
the embeddings
$\Phi_{\ee}: \Sigma \rightarrow \M$. Recalling that the identification of $\M$ and $\M_{\ee}$ is given by 
$\AA_{\ee}$ it follows that $\AA_{\ee} \circ \Phi_{\ee}$ is gauge independent, i.e.
$\AA^{(g)}_{\ee} \circ \Phi^{(g)}_{\ee} = \AA_{\ee} \circ \Phi_{\ee}$.
Under a gauge transformation $\AA_{\ee}$ transforms to $\AA^{(g)}_{\ee} = \AA_{\ee} \circ \Omega_{\ee}$, therefore
the transformation law for the embeddings is
\begin{eqnarray*}
\Phi^{(g)}_{\ee} = \Omega^{-1}_{\ee} \circ \Phi_{\ee}.
\end{eqnarray*}
Hence, the gauge transformed diffeomorphisms which make the diagram commutative (\ref{DiagramComm}) (i.e.
such that $\Phi^{(g)}_{\ee + h} = \Psi^{(g) \, \ee}_{h} \circ \Phi^{(g)}_{\ee}$) can be chosen to be
$\Psi^{(g) \, \ee}_{h} = \Omega^{-1}_{\ee + h } \circ \Psi^{\ee}_h \circ \Omega_{\ee}$, which is more
conveniently written as
$\Psi^{\ee}_{h} \circ \Omega_{\ee} = \Omega_{\ee + h } \circ \Psi^{(g) \, \ee}_h$. 
Taking the first derivative with respect to $h$ at $h=0$ we find
\begin{eqnarray}
  \vec{Z}_{\ee} (\Omega_{\ee} (x)) = \vec{R}_{\ee} (x) + \frac{\partial \Omega_{\ee}(x)}{\partial x^{\alpha}} 
Z^{(g) \, \alpha}_{\ee} (x),
\label{relee}
\end{eqnarray}
where $\vec{R}_{\ee} (x) \equiv \frac{\partial \Omega_{\ee}(x)}{\partial \ee}$. This vector
is related to $\vec{S}_{\ee}$ in (\ref{GaugeVectors}) by
\begin{eqnarray}
\vec{S}_{\ee} (x) = \left . 
\frac{\partial (\Omega_{\ee + h} \circ \Omega^{-1}_{\ee} (x))}{\partial h} \right |_{h=0} =
\vec{R}_{\ee} (\Omega^{-1}_{\ee} (x)). \label{relSR}
\end{eqnarray}
Putting $\ee=0$ in (\ref{relee}) and recalling that $\Omega_{0} = \mathbb{I}_{\M}$ the first order transformation 
$\vec{\Zone}^{(g)} = \vec{\Zone} - \vec{\Sone}$ follows. For the second order, we take an $\ee$-derivative of (\ref{relee})
at $\ee=0$. Applying directly the definitions we find
\begin{eqnarray}
\W^{\alpha} (x) + \Sone^{\beta}(x) \partial_{\beta} \Zone^{\alpha} (x) =
\left . \frac{\partial R^{\alpha}_{\ee} (x)}{\partial \ee} \right |_{\ee=0}
+ \Zone^{(g) \, \beta} (x) \partial_{\beta} \Sone^{\alpha} (x) + \W^{(g) \, \alpha} (x).
\label{intermediate.2}
\end{eqnarray}
Now, the derivative of (\ref{relSR}) at $\ee=0$ gives 
\begin{eqnarray}
\left . \frac{\partial R^{\alpha}_{\ee} (x)}{\partial \ee} \right |_{\ee=0} \equiv \left . \frac{\partial^2 
\Omega^{\alpha}_{\ee}}{\partial \ee^2} \right |_{\ee=0} =
V^{\alpha} (x) + \Sone^{\beta} (x) \partial_{\beta} \Sone^{\alpha} (x). \label{DOmegaee2}
\end{eqnarray}
Inserting this into (\ref{intermediate.2}) 
and recalling the definition of $\vec{\Stwo}$  and $\vec{\Ztwo}$ the transformation (\ref{gauge2}) follows.
$\hfill \Box$

{\bf Remark:} From (\ref{DOmegaee2}) a useful expression for $\vec{\Stwo}$ directly in terms
of derivatives of $\Omega_{\ee}$ follows (c.f. Proposition \ref{1st2nd})
\begin{eqnarray*}
\Stwo^{\alpha} (x) = \left. \frac{\partial^2 \Omega^{\alpha}_{\ee} (x)}{\partial \ee^2} \right |_{\ee=0}
+ \Gamma^{\alpha}_{\beta\gamma} (x) \Sone^{\beta} (x) \Sone^{\gamma}(x).
\end{eqnarray*}

Under spacetime gauge transformations, the tensors 
$\qper$, $\qperper$, $\kappaper$, and 
$\kappaperper$ must be gauge invariant because they are defined intrinsically on 
$\Sigma$. This provides a very strong potential check for the validity of the
expressions given in Proposition \ref{SecondPert}. The calculations required to perform the check are however
very involved  and have not been done analytically. Nevertheless, 
with the aid of a computer algebra program written in Reduce, I have checked
gauge invariance for an important number of non-trivial examples, with positive results in all cases.

As already mentioned perturbed hypersurfaces in a perturbed spacetime have two independent gauge
transformations. The first one has already been analyzed.
The second gauge freedom comes from the fact that the hypersurfaces $\Sigma_{\ee}$ embedded into
$\M_{\ee}$ must be identified with an abstract copy of $\Sigma$.
This identification entails the freedom of performing an $\ee$-dependent diffeomorphism $\chi_{\ee}$ of
$\Sigma$ before embedding this manifold into $\M$. Thus, the gauge freedom is given by the transformation $\Phi_{\ee}^{(g)} =
\Phi_{\ee} \circ \chi_{\ee}$
(recall that $\Phi_{\ee}: \Sigma  \rightarrow \M$ defines the embedded hypersurface $\Sigma_{\ee}$).
In terms of coordinates, this
gauge transformation corresponds to $\ee$-dependent coordinate changes $\hat{y}^{i} (y^j,\ee)$ of the intrinsic
coordinates in the hypersurface. It is obvious that this gauge transformation does
not affect the first and second order perturbed metrics $\Kper$, $\Kperper$. However it does affect
how points with fixed coordinates $y^{i}$ move on spacetime and therefore it affects the vectors 
$\vec{\Zone}$ and $\vec{\Ztwo}$. In order to find their gauge transformations, 
we need to define suitable first and second gauge vectors $\vec{\uone}$ and
$\vec{\utwo}$. It is clear 
from our previous discussion that they can be defined
as 
\begin{eqnarray*}
\vec{\uone} \equiv  \left . \frac{\partial \chi_{\ee}}{\partial \ee} \right |_{\ee=0}, \hspace{2cm}
\vec{\utwo}  \equiv \left. 
\frac{\partial^2 \left ( \chi_{\ee+h} \circ \chi^{-1}_{\ee} \right )}{\partial h \partial \ee} \right |_{\ee = h =0} 
+ D_{\vec{\uone}} \vec{\uone}.
\end{eqnarray*}
Being defined on $\Sigma$ they can be pushed-forward to spacetimes vectors defined on $\Sigma_0$. As before, 
we shall use the same symbol for both objects. 
\begin{proposition}
\label{3dimgauge}
Under a gauge transformation on $\Sigma$ defined by gauge vectors $\vec{\uone}$ and $\vec{\utwo}$,  
$\vec{\Zone}(p)$ and $\vec{\Ztwo}(p) $ at any point $p \in \Sigma_0$ transform as
\begin{eqnarray}
\vec{\Zone}^{(g)} (p)  =  \vec{\Zone}(p)   + \vec{\uone} (p), \label{Zoneg}\\
\vec{\Ztwo}^{(g)} (p)  = \left. \left ( \vec{\Ztwo} + \vec{\utwo}  + 2 \nabla_{\vec{\uone}} \vec{\Zone}  
- \sigma (\kappa_{ij} \uone^i \uone^j) \vec{n} \right ) \right |_{p}, \label{Ztwog}
\end{eqnarray}
where $\vec{n}$ is a unit normal, $(\vec{n}, \vec{n}) = \sigma$, and 
$\kappa_{ij}$ is the  second fundamental form of $\Sigma_0$.

\end{proposition}
{\bf Remark:} The covariant derivative in the second expression is a {\it spacetime} covariant derivative, {\it not}
a covariant derivative on $\Sigma$ push-forwarded to $\Sigma_0$. 

{\it Proof:}
A similar calculation as the one leading to (\ref{DOmegaee2}) shows that 
\begin{eqnarray}
\utwo^{i}(y) = 
\left . \frac{\partial^2 \chi^{i}_{\ee}(y)}{\partial \ee^2} \right |_{\ee=0} + \Gamma^{(3) \, i}_{jl}(y) \uone^j(y) \uone^l(y),
\label{utwo}
\end{eqnarray}
where $\Gamma^{(3) \, i}_{jl}$ are the Christoffel symbols of $(\Sigma,h)$. Writing (\ref{Z1}) with
the substitutions $\Phi_{\ee} \rightarrow \Phi^{(g)}_{\ee} = \Phi_{\ee} \circ \chi_{\ee}$,
$\vec{\Zone} \rightarrow \vec{\Zone}^{(g)}$ immediately implies  
\begin{eqnarray*}
\Zone^{(g) \, \alpha} (y) = 
\left . \frac{\partial \Phi^{\alpha}_{\ee}(y)}{\partial  \ee} \right |_{\ee=0} + 
e^{\alpha}_l(y) \left . \frac{\partial \chi_{\ee}^l(y)}{\partial \ee} \right |_{\ee=0}, \\
\end{eqnarray*}
where $e^{\alpha}_l (y)= \frac{\partial \Phi^{\alpha}(y)}{\partial y^l}$ are tangent vectors to $\Sigma_0$
and  we have used that $\chi_{\ee=0} = \mathbb{I}_{\Sigma}$. This proves (\ref{Zoneg}).
For the second derivative we find, after gauge transforming (\ref{Z2}), 
\begin{eqnarray}
\Ztwo^{(g) \, \alpha}(y) & = &
\Ztwo^{\alpha} (y) + 
\left . \frac{\partial^2 \chi^{i}_{\ee}(y)}{\partial \ee^2} \right |_{\ee=0} e^{\alpha}_i(y)+
2 u^{i} \left. \left (  e_{i}^{\beta} \nabla_{\beta} \Zone^{\alpha} \right ) \right |_y + \nonumber \\
& + &  \uone^i(y) \uone^j(y) \left (
\frac{\partial^2  \Phi^{\alpha}(y)}{\partial y^i \partial y^j} + \Gamma^{\alpha}_{\beta\gamma} (x_0(y))
e^{\beta}_i e^{\gamma}_j \right ). \label{Z2g}
\end{eqnarray}
The term in parenthesis can be rewritten as $e^{\beta}_i \nabla_{\beta} e^{\alpha}_j$. Decomposing 
this vector  into
tangent and normal components we have  $e^{\beta}_i \nabla_{\beta} e^{\alpha}_j = - \sigma \kappa_{ij} n^{\alpha}
+ \Gamma^{(3) \, l}_{ij} e_{l}^{\alpha}$, where we have used the fact that the projection on $\Sigma_0$ of
the spacetime covariant
derivative is precisely the three-dimensional covariant derivative.
Using this and (\ref{utwo}) into (\ref{Z2g}) yields the result. $\hfill \Box$

{\bf Remark:} The gauge transformation on $\Sigma$ only affects the way how points on $\Sigma$ are identified
before the $\ee$-derivative is performed.  One could think that the only effect of this should be 
transforming the vectors $\vec{\Zone}$ and $\vec{\Ztwo}$ with tangential components.
While this is clearly so for the first order vector, it 
is not true for the second variation (i.e. $\vec{\Ztwo}^{(g)} - \vec{\Ztwo}$ contains in general normal components).
The reason is that the vector $\vec{\Ztwo}$ measures essentially spacetime accelerations, and curves
fully contained within $\Sigma_0$ in general have a non-zero normal acceleration unless the hypersurface is
totally geodesic, i.e. $\kappa_{ij}=0$. Notice that in this case  the transformation (\ref{Ztwog}) does not have
normal components (provided  $\vec{\Zone}$ is tangent to $\Sigma_0$, of course).

\section{First and second order linearized matching conditions}

In this section we apply the previous results 
to perturbed matching theory.
More specifically, we will consider two spacetimes
$(\Mp,\gp)$  and   $(\Mm,\gm)$  joined  across   a  common  hypersurface
$\Sigma_0$,  i.e.  such that the  so-called matching  conditions are  satisfied on
$\Sigma_0$.  On both  regions we perturb the metric to first and second order
with $\Kper^{\pm}$ and $\Kperper^{\pm}$ respectively.  Our aim is to obtain
necessary and sufficient conditions on $\Sigma_0$  such  that  the matching
conditions are also satisfied in a perturbed sense (i.e. to first and second order).

Let us start with a  brief discussion of the matching conditions. 
Matching theory between  spacetimes deals with two $C^2$  spacetimes
$(\Mp,\gpm)$  with  boundary.  The  respective  boundaries  are  $C^3$
hypersurfaces   of    $\Mpm$   which   are    called   {\it   matching
hypersurfaces}.  We will  denote  them by  $\Sigma^{\pm}$. Although  a
fully successful theory  can be developed for boundary  of an arbitrary
causal character \cite{MarsSenovilla93}, for simplicity we will 
concentrate here in the case where
 both $\Sigma^{\pm}$  are  either timelike  or
spacelike  everywhere.  The  matching  theory  asserts  that  a  $C^0$
spacetime $(\Mt,\gt)$  can be  constructed by joining  $(\Mp,\gp)$ and
$(\Mm,\gm)$ if
and  only  if there  exists  a  diffeomorphism  $\varphi :  \Sigma^{+}
\rightarrow  \Sigma^{-}$ which  is  an isometry  with  respect to  the
induced  metrics  $h^{\pm}$ on  each  boundary. This is equivalent to
introducing an   abstract
$(n-1)$-dimensional  $C^3$   manifold  $\Sigma$,  and   demanding  the
existence of  two embeddings $\Phi^{\pm}:  \Sigma \rightarrow \M^{\pm}$
such that (i)  $\Phi^{\pm} (\Sigma) = \Sigma^{\pm}  \subset \M^{\pm}$
and (ii)  that    the   two   induced    metrics   $h^{\pm}   \equiv
{\Phi^{\pm}}^{\star}(  \gpm )$  on  $\Sigma$ coincide,  i.e. $h^{+}  =
h^{-}$.   Condition (ii)  is   called  {\it  first  set  of  matching
conditions}.
Furthermore, the Riemann tensor in the joined spacetime  $(\Mt,\gt)$  is
free of Dirac delta distributions if and only if the second fundamental forms
on $ \Sigma$ coincide, i.e.
\begin{eqnarray}
{\Phi^{+}}^{\star}   \left  (   \nabla^{+}  \bm{n}^{+}   \right   )  =
{\Phi^{-}}^{\star} \left ( \nabla^{-} \bm{n}^{-} \right ),
\label{SecondConditions}
\end{eqnarray}
where $\pm$ always means that the objects are calculated from the $\M^{\pm}$ side. 
The unit normal $\bm{n}^{\pm}$ to $\Sigma^{\pm}$ are chosen to have
the  same  relative  orientation  after  the  matching,  i.e.   either
$\vec{n}^{+}$   points    outwards   and   $\vec{n}^{-}$    inwards   or
vice versa (``inwards'' and ``outwards'' are well-defined
concepts for  vector fields 
on the  boundary  of  a manifold-with-boundary).  
Notice  that the matching
conditions are fully covariant both with respect to the spacetimes 
$\M^{\pm}$  and with  respect  to  the matching  hypersurface
$\Sigma$. This means that any local coordinate system in $\Mp$ and any
coordinate system in  $\Mm$ are equally valid to  impose the matching
conditions.  Moreover  any coordinate system can be used  in the abstract
manifold  $\Sigma$. 

We want to perturb the spacetimes with boundary $(\M^{\pm},g^{\pm})$ 
and analyze under which conditions the matching conditions are satisfied perturbatively (assuming
the background spacetimes do match across their boundary). First of all, we
need to consider how
to define perturbations for manifolds with boundary. As mentioned above, 
in important ingredient in perturbation theory is the need
to identify the different manifolds $\M_{\ee}$ with each other.
Now, our manifolds $\M^{+}_{\ee}$ have boundary
(we concentrate on the ``+'' side; similar considerations hold for $\M^{-}_{\ee}$). 
If we imagine them 
as subsets of larger
manifolds without boundary $\tilde{\M}_{\ee}$  and identify these, 
it is clear that, generically, the identification will not transform the 
boundaries $\Sigma_{\ee}$ among themselves.
We could insist that the identification preserves the boundaries,
but only at the expense of restricting strongly the 
gauge freedom (at least near
the boundary). This may be quite inconvenient for other purposes. Thus, it is 
better to let
the boundary ``move'' freely
in the identification. From the point of view of the family of manifolds 
with boundary this means that,
strictly speaking, we are not
taking diffeomorphisms between them.
They 
are diffeomorphisms except in some closed neighbourhood of the boundary $\supIe$, for 
each $\epsilon$. Thus,
in strict terms,
we cannot talk of a background manifold $\mmIo$ with boundary $\supIo$ and a 
family of metrics $g_{\ee}$ defined on it.
Nevertheless, in perturbation theory we only care about derivatives 
at $\epsilon=0$ of the 
$\epsilon$-family 
and this can be consistently defined
on points at the boundary $\supIo$ by taking one-sided derivatives (i.e. 
restricting the variations to positive
or to negative $\epsilon$ depending on the point of the boundary we are considering). This 
allows one to define, similarly as in the case without boundary, a
background manifold with boundary $\M^{+}$ and two symmetric tensors $\Kper^{+}$ 
and $\Kperper^{+}$ defined everywhere up to
and including the boundary, so that we have a proper perturbation theory
up to second order (or higher order if desired).
Having this in mind, we will abuse notation and still talk about
diffeomorphisms between different $\mmIe$. This also allows us to talk about
how  the hypersurfaces $\Sigma_{\ee}$ move on the background and therefore introduce vectors 
$\Zone^{+}$ and $\Ztwo^{+}$ on the unperturbed boundary, exactly as we did in Section \ref{Z1Z2}.

With this particularity in mind, it now easy to write down the perturbed matching conditions. Indeed, for each
$\ee$ the matching conditions demand the equality of the first and second fundamental forms on each side, i.e.
$h_{\ee}^{+}  = h^{-}_{\ee}$, $\kappa^{+}_{\ee} = \kappa^{-}_{\ee}$. This tensors are all defined
on the abstract hypersurface
$\Sigma$  and therefore can be compared with each other (and differentiated with respect to $\ee$). 
It is also clear that the matching conditions
will be satisfied in a perturbed sense if and only if the first and second derivatives of the first and
second fundamental forms coincide from both sides. 
Using the explicit form of these derivatives found in previous sections 
we have a practical method of determining whether perturbations of a spacetime constructed by
joining two regions across $\Sigma_0$ can be matched across this hypersurface. We state this result
in the form of a theorem
\begin{theorem}
\label{PertMatch}
Let $(\M,g)$ be a spacetime constructed by joining two spacetimes with boundary $(\M^{+},g^{+})$ and 
$(\M^{-},g^{-})$ across their corresponding boundaries $\Sigma^{\pm}$.
Let $\Sigma$ be an abstract copy of $\Sigma^{+}$ 
and $\Phi^{\pm} : \Sigma \rightarrow \M^{\pm}$ be the embeddings defining the background matching.
Let also $\Kper^{\pm}$ and $\Kperper^{\pm}$ be
first and second order metric perturbations in $\M^{\pm}$. 

The first
order perturbed (i.e. linearized) matching conditions are fulfilled if and only if there exist two scalars
$\Qone^{\pm}$ and two vectors $\vec{\Tone}^{\pm}$ on $\Sigma$ for which
\begin{eqnarray*}
\qper^{+}_{ij} = \qper^{-}_{ij}, \hspace{2cm}
\kappaper^{+}_{ij} = \kappaper^{-}_ {ij},
\end{eqnarray*}
holds, where $\qper^{\pm}$, $\kappaper^{\pm}_{ij}$ are given in Proposition \ref{FirstPert}
after the substitution
$\Qone \rightarrow \Qone^{\pm}$,
$\vec{\Tone} \rightarrow \vec{\Tone}^{\pm}$, $g \rightarrow g^{\pm}$, $\Kper \rightarrow \Kper^{\pm}$ and $e^{\alpha}_{i} \rightarrow
e^{\alpha \, \pm}_{i}$.

The second order perturbed matching conditions are satisfied if and only if there exist two
 scalars $\Qtwo^{\pm}$ and
two vector fields $\vec{\Ttwo}^{\pm}$ on $\Sigma$ such that 
\begin{eqnarray*}
\qperper^{+}_{ij} = \qperper^{-}_{ij}, \hspace{2cm}
\kappaperper^{+}_{ij} = \kappaperper^{-}_ {ij},
\end{eqnarray*}
where these objects are obtained from (\ref{qperper})-(\ref{kappaperper}) after similar substitutions.
\end{theorem}
{\bf Remark:} It is important to stress the fact that satisfying the perturbed matching
conditions require the {\it existence} of vector fields $\vec{\Zone}^{\pm}$ and $\vec{\Ztwo}^{\pm}$
such the equations above are satisfied. These vector fields are not known a priori. Moreover they {\it need}
not be the same vector on both sides. This is obvious from the fact that these vectors are gauge dependent
and the gauge may be chosen differently in the different regions $\M^{\pm}$ (actually  one can not
even compare the two gauges, in general). The gauge can always be chosen so that these vectors coincide but this may
not be the most convenient choice. 
Linearized matching conditions have often been analyzed by using specific gauges where
the vectors $\vec{\Zone}$ and $\vec{\Ztwo}$ take simpler forms. A common choice is to use Gauss coordinates adapted
to the matching hypersurfaces for all $\ee$ (this obviously makes $\vec{\Zone}^{+} = \vec{\Zone}^{-}=0$) and then
transform to the desired gauge. This is the approach taken in \cite{Kind93} for instance. In spherical symmetry, the
linearized matching conditions in arbitrary gauge was first studied in \cite{Gerlach79.1}, \cite{Gerlach79.2}
and completed in \cite{MartinGarcia}. The general linearized matching conditions in an arbitrary background were
first presented by Mukohyama \cite{Mukohyama00}. In this paper a vector field $\vec{Z}$ was introduced describing the
perturbation  of the matching hypersurface to first order. However, this vector was assumed to  be the same
in both sides $\M^{+}$ and $\M^{-}$. As I have already stressed this need not be case and the fully general
perturbed matching conditions require the use of two vectors $\vec{\Zone}^+$ and $\vec{\Zone}^{-}$ (and two more vectors
to second order). Notice also that the gauge freedom within $\Sigma$  (see Proposition \ref{3dimgauge})
allows us to choose the tangential components of $\vec{\Zone}$ and $\vec{\Ztwo}$ in any way we want, but
only on one of the sides, i.e. either on  $\M^{+}$ or on $\M^{-}$. Once a choice on one side has been made, the other side must be left free
and determined by the matching conditions (if they happen to be consistent). This is similar to the fact that when solving
a matching problem one not only looks for a pair of  matching hypersurfaces with suitable properties, but {\it also} for a
specific pair of embeddings on each side, i.e. a way of identifying the two hypersurfaces pointwise.

In this theorem, only non-null hypersurfaces are considered. This 
is because the classical
Darmois matching conditions (discussed above) are {\it not} adequate 
for hypersurfaces with null points. In that case the
continuity of the second fundamental form does not ensure the absence
of distributional parts in the Riemann tensor.
The matching conditions for null hypersurfaces where first
discussed by Clarke and Dray \cite{ClarkeDray87} and later extended to
hypersurfaces of arbitrary causal character (including a changing one)
in \cite{MarsSenovilla93}. They involve the continuity of a tensor that generalizes
the second fundamental form. In order to find the perturbed matching conditions in this case
we would need to find how this new tensor is perturbed to second order. This issue
is of interest and should be studied. The methods described in the present paper
are useful to find perturbations of any geometric tensor defined on a hypersurface and 
therefore are applicable to this situation too. The calculations
for hypersurfaces with null points are probably more difficult but still manageable.

\section*{Acknowledgements}

I wish to thank Ra\"ul Vera for a careful reading of the manuscript and for
interesting suggestions and discussions. Thanks also to Malcolm A.H. MacCallum for
discussions during a stay in the School of Mathematical Sciences, Queen Mary College,  University
of London,  where part of this work was carried out. Financial support for this visit
under the EPSRC grant GR/R53685/01 is acknowledged. Part of this work has been done under the project 
BFME2003-02121 of the Spanish Ministerio de Educaci\'on y Tecnolog\'{\i}a.

\section{Appendix A: Proof of Proposition \ref{SecondPert}}

{\it Proof:} 
Let us start with $\qperper$. From Lemma \ref{FirstSecondOrder} we find
\begin{eqnarray}
\left . \partial_{\ee} \partial_{\ee} h_{\ee} \right |_{\ee=0} = \Phi^{\star} \left ( \Kperper 
+ 2 \pounds_{\vec{\Zone}} \Kper + \pounds_{\vec{\W}} g + \pounds_{\vec{\Zone}} \pounds_{\vec{\Zone}} g \right ).
\label{Pullbackqperper}
\end{eqnarray}
Applying (\ref{LieZZBbis}) with $B \rightarrow g$ and using (\ref{LieF1F2g}) with $F_1 = F_2 = \Qone$ we readily obtain
\begin{eqnarray*}
\pounds_{\vec{\Zone}} \pounds_{\vec{\Zone}} g_{\alpha\beta} = \pounds_{\nabla_{\vec{\Zone}}\vec{\Zone}} g_{\alpha\beta}
+ 2 \pounds_{\vec{\Tone}} \pounds_{\vec{\Zone}} g_{\alpha\beta} - \pounds_{\vec{\Tone}} \pounds_{\vec{\Tone}} g_{\alpha\beta}
- \pounds_{\vec{\Done}} g_{\alpha\beta} + 
 2 \left ( \frac{}{} \sigma \Tone^{\mu} \Tone^{\nu} \kappa_{\mu\nu} \right .   \\ \left . 
- 2 \vec{\Tone} ( \Qone ) \right ) 
\kappa_{\alpha\beta} 
+ 2 \Qone^2 \left (  -n^{\mu} n^{\nu} R_{\alpha\mu\beta\nu} + \kappa_{\alpha\mu}
\kappa_{\beta}^{\,\,\mu} \right ) + 2 \sigma D_{\alpha} \Qone D_{\beta} \Qone + n_{\alpha} \P_{\beta} + n_{\beta} \P_{\alpha}.
\end{eqnarray*}
Here and in the following $\P_{\alpha}$, $\P_{\alpha\beta}, \cdots$
 stands for expressions whose explicit form does not concern us. Notice that
its  meaning may be different even in different parts of the same formula.

Substituting into (\ref{Pullbackqperper}) we observe that a term $2\pounds_{\vec{\Tone}} (
\Kper + \pounds_{\vec{\Zone}} g )$ appears. From Proposition \ref{FirstPert},
the pull-back of this term is just $2 \pounds_{\vec{\Tone}} \qper$. Recalling that 
 $\vec{\Ztwo} \equiv \vec{W} + 
\nabla_{\vec{\Zone}} \vec{\Zone}$ and the decomposition $\vec{\Ztwo}= \Qtwo \vec{n} + \vec{\Ttwo}$
yields the first two terms in (\ref{qperper}).
Only  $2\pounds_{\Qone \vec{n}} \Kper_{\alpha\beta}$ remains to be analyzed. This is dealt with using
Lemma \ref{LieS} with $\vec{X} \rightarrow \Qone  \vec{n}$ which gives
\begin{eqnarray*}
2 \pounds_{\Qone \vec{n}} \Kper_{\alpha\beta} = 
- 4 \Qone n_{\mu} \Sper^{\mu}_{\alpha\beta} + 2 \pounds_{\Qone \vec{\Kpernortan}}
g_{\alpha\beta} + 4 \sigma \Qone \Kpernornor \kappa_{\alpha\beta} + n_{\alpha} \P_{\beta} + n_{\beta} \P_{\alpha},
\end{eqnarray*}
and expression (\ref{qperper}) follows directly.
Let us next consider $\kappaperper$, which involves  the longest and most difficult calculation. 
Applying Lemma \ref{FirstSecondOrder} to $\kappa_{\ee}$ we find
\begin{eqnarray}
\left . 2 \partial_{\ee} \partial_{\ee} \kappa_{\ee}  \right |_{\ee=0} = \Phi^{\star} \left ( \pounds_{\vec{n}_2} g
+ 2 \pounds_{\vec{n}_1} \Kper +  \pounds_{\vec{n}} \Kperper + 2 \pounds_{\vec{\Zone}} \pounds_{\vec{n}_1} g + 
2 \pounds_{\vec{\Zone}} \pounds_{\vec{n}} \Kper + \right . \nonumber \\
\left . + \pounds_{\vec{\W}} \pounds_{\vec{n}} g + \pounds_{\vec{\Zone}} 
\pounds_{\vec{\Zone}} \pounds_{\vec{n}} g \right ).
\label{extrinperper}
\end{eqnarray}
In Proposition \ref{FirstPert} we evaluated $\partial_{\ee} \kappa_{\ee}
|_{\ee=0}$, which required calculating the pull-back on $\Sigma$ of
$M \equiv
\pounds_{\vec{\Zone}} \pounds_{\vec{n}} g +
\pounds_{\vec{n}_1} g + \pounds_{\vec{n}}
\Kper$ (see (\ref{kappaper.1})). We want to identify in (\ref{extrinperper}) terms giving
the Lie derivative of $M$ along
$\vec{\Tone}$. Adding and subtracting
$2 \pounds_{\vec{\Tone}} \pounds_{\vec{\Zone}} \pounds_{\vec{n}} g$
we can write
\begin{eqnarray}
\left . 2 \partial_{\ee} \partial_{\ee} \kappa_{\ee}  \right |_{\ee=0}
= \Phi^{\star} \left ( 
\pounds_{\vec{n}_2} g
+ 2 \pounds_{\vec{n}_1} \Kper +  \pounds_{\vec{n}} \Kperper + 2
\pounds_{ \Qone \vec{n}} \pounds_{\vec{n}_1} g + 
2 \pounds_{\Qone \vec{n} } \pounds_{\vec{n}} \Kper + \right
. \nonumber \\
\left . + \pounds_{\vec{\W}} \pounds_{\vec{n}} g + \pounds_{\vec{\Zone}} 
\pounds_{\vec{\Zone}} \pounds_{\vec{n}} g 
- 2 \pounds_{\vec{\Tone}} \pounds_{\vec{\Zone}} \pounds_{\vec{n}} g
+ 2 \pounds_{\vec{\Tone}} M \right ).
\label{extrinperperbis}
\end{eqnarray}
Let us deal with the different terms in this expression, starting
with $\pounds_{\vec{\none}} \Kper$. A direct application of (\ref{LieQn})
and the explicit expression
for $\vec{\none}$ (\ref{none}) yield, after using (\ref{LieKperSper}),
\begin{eqnarray}
\pounds_{\vec{\none}} \Kper_{\alpha\beta} = - \Kpernornor^2
\kappa_{\alpha\beta}
-\frac{\sigma}{2}  \Kpernornor \pounds_{\vec{\Kpernortan}}
g_{\alpha\beta}
+ \sigma \Kpernornor n_{\mu} \Sper^{\mu}_{\alpha\beta}
- \frac{\sigma}{2}  \left ( \Kpernortan_{\alpha} D_{\beta} \Kpernornor
+ \Kpernortan_{\beta} D_{\alpha} \Kpernornor \right ) - \nonumber \\
- \pounds_{\vec{\Kpernortan} + \Qone \vec{a} + \sigma \grad 
  \Qone} \Kper_{\alpha\beta} + n_{\alpha} \P_{\beta} + n_{\beta}
\P_{\alpha}. \label{Lien1Kper}
\end{eqnarray}
Next, we analize the combination
$2 \pounds_{\Qone \vec{n}} \pounds_{\vec{\none}} g+
2 \pounds_{\Qone \vec{n}} \pounds_{\vec{n}} \Kper$. Using expression (\ref{none}) and
the first equality in (\ref{LieKperSper}) we get
\begin{eqnarray*}
2 \pounds_{\Qone \vec{n}} \pounds_{\vec{\none}} g_{\alpha\beta}
+2 \pounds_{\Qone \vec{n}} \pounds_{\vec{n}} \Kper_{\alpha\beta}=
\sigma \pounds_{\Qone \vec{n}} \pounds_{\Kpernornor \vec{n}} g_{\alpha\beta}
- 2 \pounds_{\Qone \vec{n}} \pounds_{\Qone \vec{a} + \sigma
  \grad Q} g_{\alpha\beta} 
- 4 \pounds_{\Qone \vec{n}} \left 
( n_{\mu} \Sper^{\mu}_{\alpha\beta} \right ).
\end{eqnarray*}
Using now identity (\ref{LieF1F2g}) with $F_1 = \Qone$, $F_2 =
\Kpernornor$ and writing the Lie derivative of the last term explicitly 
using covariant derivatives yields
\begin{eqnarray}
2 \pounds_{\Qone \vec{n}} \pounds_{\vec{\none}} g_{\alpha\beta}
+2 \pounds_{\Qone \vec{n}} \pounds_{\vec{n}} \Kper_{\alpha\beta} =
\sigma \pounds_{\Qone \Kpernornor \vec{a}} g_{\alpha\beta} 
\nonumber \label{LieQn1g} \\ 
+ 2 \sigma  \Qone \Kpernornor
\left ( - n^{\mu} n^{\nu} R_{\alpha\mu\beta\nu} 
+ \kappa_{\alpha\mu} \kappa_{\beta}^{\,\,\,\mu} \right ) 
+ \left ( \nabla_{\alpha} \Qone \nabla_{\beta} \Kpernornor +
\nabla_{\alpha} \Kpernornor \nabla_{\beta} \Qone \right ) + \\
+ 2 \sigma \kappa_{\alpha\beta} \Qone \vec{n} (\Kpernornor )
- 2 \pounds_{\Qone \vec{n}} \pounds_{\Qone
\vec{a} + \sigma \grad \Qone} g_{\alpha\beta} 
- 4 \Qone a_{\mu} \Sper^{\mu}_{\alpha\beta}
- 4 \Qone n_{\mu} n^{\nu} \nabla_{\nu} \Sper^{\mu}_{\alpha\beta} - \nonumber \\
- 4 n_{\mu} n^{\nu} \Sper^{\mu}_{\alpha\nu} \nabla_{\beta} \Qone
- 4 n_{\mu} n^{\nu} \Sper^{\mu}_{\beta\nu} \nabla_{\alpha} \Qone
- 4 n_{\mu} \Sper^{\mu}_{\alpha\nu} \kappa^{\nu}_{\, \, \beta}
- 4 n_{\mu} \Sper^{\mu}_{\beta\nu} \kappa^{\nu}_{\, \, \alpha}
+ n_{\alpha} \P_{\beta} + n_{\beta} \P_{\alpha}. 
\nonumber
\end{eqnarray}
Next we analyze the terms involving third derivatives 
in (\ref{extrinperperbis}). Lemma \ref{LieZZ} implies
\begin{eqnarray*}
\pounds_{\vec{\W}} \pounds_{\vec{n}} g + \pounds_{\vec{\Zone}} 
\pounds_{\vec{\Zone}} \pounds_{\vec{n}} g 
- 2 \pounds_{\vec{\Tone}} \pounds_{\vec{\Zone}} \pounds_{\vec{n}} g = \\
= \pounds_{\vec{\Ztwo}} \pounds_{\vec{n}} g -
\pounds_{\vec{\Tone}} \pounds_{\vec{\Tone}} \pounds_{\vec{n}} g -
\pounds_{\Cone \vec{n} + \vec{\Done}} \pounds_{\vec{n}} g
- \pounds_{\Qone^2 \vec{a}} \pounds_{\vec{n}} g +
\pounds_{\Qone \vec{n}} \pounds_{\Qone \vec{n}} \pounds_{\vec{n}} g.
\end{eqnarray*}
Except for the last term
$\pounds_{\Qone \vec{n}} \pounds_{\Qone \vec{n}} \pounds_{\vec{n}} g$,
the pull-back of the right-hand side is easily obtained from 
Lemma \ref{embed} and  (\ref{LieF1F2g}). For the last term we have, directly
from (\ref{LieQ1g}),
\begin{eqnarray}
\pounds_{\Qone \vec{n}} \pounds_{\Qone \vec{n}} \pounds_{\vec{n}} g
= 
\pounds_{\Qone \vec{n}} \pounds_{\Qone \vec{a}} g
+ 2 \Qone \vec{n} (\Qone ) \left ( 
- n^{\mu} n^{\nu}R_{\alpha\mu\beta\nu} 
+ \kappa_{\alpha\mu} \kappa_{\beta}^{\,\,\,\mu} \right ) 
+ 2 \Qone \pounds_{\Qone \vec{n}} \left (
\kappa_{\alpha\mu} \kappa_{\beta}^{\,\,\,\mu} - \right . \nonumber \\ \left .  
 n^{\mu} n^{\nu}R_{\alpha\mu\beta\nu} 
\right )
+ 2 \sigma \Qone \left ( \frac{}{}
a^{\mu} \kappa_{\alpha\mu} \left (
\Qone a_{\beta} + \sigma D_{\beta} \Qone \right )+ 
a^{\mu} \kappa_{\beta\mu} \left (
\Qone a_{\alpha} + \sigma D_{\alpha} \Qone \right )
\right ) + 
n_{\alpha} \P_{\beta} + n_{\beta} \P_{\alpha}. 
\label{LieQnQnng.1}
\end{eqnarray}
In order to elaborate this expression further we expand the Lie derivative in the third summand
in terms of covariant derivatives. We get
\begin{eqnarray}
\label{LieRiemann}
n^{\mu} n^{\nu} \pounds_{\Qone \vec{n}} R_{\alpha\mu\beta\nu} =
\Qone n^{\mu} n^{\nu} n^{\delta} \nabla_{\delta} R_{\alpha\mu\beta\nu} 
+ 2 \vec{n} (\Qone) n^{\mu} n^{\nu} R_{\alpha\mu\beta\nu} + \nonumber \\
\Qone \left ( a^{\mu}n^{\nu}R_{\alpha\beta\mu\nu} + 
n^{\mu} a^{\nu} R_{\alpha\beta\mu\nu} + 
\kappa_{\alpha}^{\, \,\delta} R_{\delta\mu\beta\nu} n^{\mu} n^{\nu} 
+ \kappa_{\beta}^{\, \,  \delta} R_{\delta\mu\alpha\nu} n^{\mu} n^{\nu} 
\right ) + n_{\alpha}\P_{\beta} + n_{\beta} \P_{\alpha}.
\end{eqnarray}
For the Lie derivative of $\kappa_{\alpha\beta}$ along $\Qone \vec{n}$ we have
\begin{eqnarray}
\pounds_{\Qone \vec{n}} \kappa_{\alpha\beta} =
- \sigma \Qone a_{\alpha} a_{\beta} + \Qone \left ( 
- n^{\mu} n^{\nu}R_{\alpha\mu\beta\nu} 
+ \kappa_{\alpha\mu} \kappa_{\beta}^{\,\,\,\mu} \right )
+ \frac{\Qone}{2}  h^{\mu}_{\alpha} h^{\nu}_{\beta} \pounds_{\vec{a}} g_{\mu\nu},
\label{Liekappa}
\end{eqnarray}
which follows from
$\pounds_{\Qone \vec{n}} h^{\alpha}_{\beta} = - \sigma ( \Qone a_{\beta} + \sigma D_{\beta} \Qone )
n^{\alpha}$ and $\kappa_{\alpha\beta} = \frac{1}{2} h^{\mu}_{\alpha} h^{\nu}_{\beta} \pounds_{\Qone \vec{n}} g_{\mu
\nu}$ after applying (\ref{LieQ1g}). Inserting (\ref{LieRiemann}) and (\ref{Liekappa}) into
(\ref{LieQnQnng.1}) the following expression is found
\begin{eqnarray}
\label{LieQnQnng.2}
\pounds_{\Qone \vec{n}} \pounds_{\Qone \vec{n}} \pounds_{\vec{n}} g_{\alpha\beta} =
\pounds_{\Qone \vec{n}} \pounds_{\Qone \vec{a}} g_{\alpha\beta}
+ 2 \Qone \vec{n} (\Qone ) \left ( 
- n^{\mu} n^{\nu}R_{\alpha\mu\beta\nu} 
+ \kappa_{\alpha\mu} \kappa_{\beta}^{\,\,\,\mu} \right ) \nonumber \\
- 2 \Qone^2 \left ( n^{\mu} n^{\nu} n^{\delta} \nabla_{\delta} R_{\alpha\mu\beta\nu} 
+ a^{\mu}n^{\nu}R_{\alpha\mu\beta\nu} + 
n^{\mu} a^{\nu} R_{\alpha\mu\beta\nu} + 
2 \kappa_{\alpha}^{\, \,\delta} R_{\delta\mu\beta\nu} n^{\mu} n^{\nu} 
+ 2 \kappa_{\beta}^{\, \,  \delta} R_{\delta\mu\alpha\nu} n^{\mu} n^{\nu} 
\right ) \nonumber \\
+ 2 \Qone a^{\mu} \kappa_{\mu\beta} D_{\alpha}\Qone 
+ 2 \Qone a^{\mu} \kappa_{\mu\alpha} D_{\beta}\Qone  
 + \Qone^2 \left ( \kappa^{\,\, \mu}_{\beta} \pounds_{\vec{a}} g_{\alpha\mu} 
+ \kappa^{\,\, \mu}_{\alpha}  \pounds_{\vec{a}} g_{\beta\mu}\right ) + 
n_{\alpha}\P_{\beta} + n_{\beta} \P_{\alpha}.
\end{eqnarray}
It only remains to calculate the first term in (\ref{extrinperperbis}), i.e. to
find the vector $\vec{\ntwo} = \partial_{\ee} \partial_{\ee} \vec{n}_{\ee} |_{\ee=0}$. 
The calculation is somewhat long and will be given in Appendix B. The result is
\begin{eqnarray}
\ntwo^{\alpha} = n^{\alpha} \left ( - \frac{1}{2} \sigma \Kperpernornor + \frac{3}{4} \Kpernornor^2 + \sigma 
\Kpernortan_{\mu} 
\Kpernortan^{\mu} \right )
- \Kperpernortan^{\alpha} 
- \left ( \Qtwo a^{\alpha} + \sigma D^{\alpha} \Qtwo \right ) 
+ \sigma \Kpernornor \Kpernortan^{\alpha} \nonumber \\ 
+ \left ( \Cone a^{\alpha} + \sigma D^{\alpha} \Cone \right ) 
- \left ( \vec{n} (\Qone) + \sigma \Kpernornor \right ) 
\left ( \Qone a^{\alpha} + \sigma D^{\alpha} \Qone \right ) 
+ 2 \Kpertan^{\alpha \beta} \left ( \Kpernortan_{\beta}
+ \Qone a_{\beta} + \sigma D_{\beta} \Qone \right ) \nonumber \\ +  
\left [ \Qone \vec{n}, \Qone \vec{a} + \sigma \gradL \Qone  \right ]^{\alpha}
+ 2 \Qone^2 a^{\mu} \kappa^{\alpha}_{\, \, \mu} + 2 \sigma \Qone 
\kappa^{\alpha}_{\, \, \mu} D^{\mu} \Qone.
\label{n2.2}
\end{eqnarray}
Our interest is to calculate $\pounds_{\vec{\ntwo}} g_{\mu\nu}$. Only the term
involving $2 \Kpertan^{\alpha \beta}  ( \Kpernortan_{\beta}
+ \Qone a_{\beta} + \sigma D_{\beta} \Qone )$ requires further analysis. Applying
Lemma \ref{LieS} and the fact that $\Sper^{\mu}_{\alpha\beta} = S (\Kpertan)^{\mu}_{\alpha\beta}
+ \sigma \tau^{\mu} \kappa_{\alpha\beta}  + n^{\mu} \P_{\alpha\beta} + 
n_{\alpha} \P^{\,\, \mu}_{\beta} + 
n_{\beta} \P^{\,\, \mu}_{\alpha}$, which follows directly from the definition of $\Sper$ and the decomposition
(\ref{decomKper}), we get
\begin{eqnarray*}
\pounds_{2 \Kpertan^{\mu \nu} \left ( \Kpernortan_{\nu}
+ \Qone a_{\nu} + \sigma D_{\nu} \Qone \right )} g_{\alpha \beta} =
2 \pounds_{\vec{\Kpernortan} + \Qone \vec{a} + 
\sigma \grad \Qone} \Kpertan_{\alpha\beta}+ \hspace{4cm} \\ 
+  4 \left (\Kpernortan_{\mu} + \Qone a_{\mu}
+ \sigma D_{\mu} \Qone \right ) \left ( \Sper^{\mu}_{\alpha\beta} - \sigma^{\mu} \kappa_{\alpha\beta} \right )
+ n_{\alpha} \P_{\beta} + n_{\beta} \P_{\alpha}.
\end{eqnarray*}
We are now in a position where all terms in (\ref{extrinperperbis}) can be collected.
Several obvious cancellations happen which
will not be described in any detail. More subtle is the use of the following identity
\begin{eqnarray*}
- \pounds_{\Qone \vec{a}} \pounds_{\Qone \vec{n}} g_{\alpha\beta} -
\pounds_{\Qone^2 \vec{a}} \pounds_{\vec{n}} g_{\alpha\beta}
+ \Qone^2 \kappa^{\mu}_{\alpha} \pounds_{\vec{a}} g_{\beta\mu} 
+ \Qone^2 \kappa^{\mu}_{\beta} \pounds_{\vec{a}} g_{\alpha\mu}
+ 2 \Qone \left ( a^{\mu} \kappa_{\mu \beta} D_{\alpha} \Qone
+  a^{\mu} \kappa_{\mu \alpha} D_{\beta} \Qone \right ) \\
- 2 \Qone^2 \left ( n^{\mu} a^{\nu} R_{\alpha\mu\beta\nu} + a^{\mu} n^{\nu} R_{\alpha\mu\beta\nu} \right )
+ 2 \Qone \vec{a} ( \Qone ) \kappa_{\alpha\beta} + 2 \pounds_{\Qone^2 a^{\nu} \kappa^{\mu}_{\, \nu} } g_{\alpha\beta} = 
n_{\alpha} \P_{\beta} + n_{\beta} \P_{\alpha}.
\end{eqnarray*}
This expression follows by direct calculation using the Codazzi identity written 
in the spacetime form 
$D_{\alpha} \kappa_{\mu\nu} - D_{\mu} \kappa_{\alpha\nu} =  n^{\delta} R_{\sigma \delta \beta\rho} 
h^{\sigma}_{\nu} h^{\beta}_{\alpha} h^{\rho}_{\mu}$ and the fact that $D_{\alpha} a_{\beta} -
D_{\beta} a_{\alpha} =0$, which is a direct consequence of the definition of acceleration in our hypersurface
orthogonal case.

Finally, in order to arrive at the final expression,
two more ingredients are required. The first one is
\begin{eqnarray*}
\left [ \Qone \vec{n}, \gradL \Qone \right ]^{\alpha} & = & 
\Qone D^{\alpha} (\vec{n} (\Qone)) - \sigma \Qone \vec{n} (\Qone) a^{\alpha} - 2 \Qone 
\kappa^{\alpha}_{\mu} D^{\mu} \Qone \\
& & - \left ( D_{\mu} \Qone D^{\mu} \Qone + \sigma \Qone \vec{a} (\Qone ) \right ) n^{\alpha},
\end{eqnarray*}
which is checked directly. The second one is $\vec{n} (\Kpernornor ) = 2 \Kpernortan_{\alpha}a^{\alpha}
+ 2 n_{\mu} n^{\rho} n^{\delta}\Sper^{\mu}_{\rho\delta}$, which is immediate.
Using also the explicit expression for $\Cone$
and $\vec{\Done}$ in (\ref{Cone}) and collecting all terms we find
(\ref{kappaperper}). $\hfill \Box$

\section{Appendix B: Calculation of $\vec{\ntwo}$}

In this appendix we find an explicit expression for $\partial_{\ee} \partial_{\ee} \vec{n}_{\ee}$.

\begin{lemma}
With the same notation and conventions as in Proposition \ref{SecondPert} we have
\begin{eqnarray}
\ntwo^{\alpha} = 
n^{\alpha} \left ( - \frac{1}{2} \sigma \Kperpernornor + \frac{3}{4} \Kpernornor^2 + \sigma 
\Kpernortan_{\mu} 
\Kpernortan^{\mu}  \right ) 
- \Kperpernortan^{\alpha} 
- \left ( \Qtwo a^{\alpha} + \sigma D^{\alpha} \Qtwo \right ) 
+ \sigma \Kpernornor \Kpernortan^{\alpha} \nonumber \\
+ \left ( \Cone a^{\alpha} + \sigma D^{\alpha} \Cone \right ) 
- \left ( \vec{n} (\Qone) + \sigma \Kpernornor \right ) 
\left ( \Qone a^{\alpha} + \sigma D^{\alpha} \Qone \right ) 
+ 2 \Kpertan^{\alpha \beta} \left ( \Kpernortan_{\beta}
+ \Qone a_{\beta} + \sigma D_{\beta} \Qone \right )  \nonumber \\   +  
\left [ \frac{}{} \Qone \vec{n}, \Qone \vec{a} + \sigma \gradL  \Qone  \right ]^{\alpha}
+ 2 \Qone^2 a^{\mu} \kappa^{\alpha}_{\, \, \mu} + 2 \sigma \Qone 
\kappa^{\alpha}_{\, \, \mu} D^{\mu} \Qone.
\label{n2.2.bis}
\end{eqnarray}
\end{lemma}
{\it Proof:}
We proceed as we did for $\vec{\none}$, i.e. we first determine the normal component of $\vec{\ntwo}$
and then its tangential part. 
Taking the second $\ee$ derivative of
$( \vec{n}_{\ee} , \vec{n}_{\ee} )_{\gee}= \sigma$ and evaluating at $\ee=0$ we find
\begin{eqnarray*}
2 \ntwo^{\mu} n_{\mu}  + 2 \none^{\mu} \none_{\mu} + 4 \none^{\mu}
n^{\nu} \Kper_{\mu\nu} + n^{\mu} n^{\nu} \Kperper_{\mu\nu} = 0, 
\end{eqnarray*}
which after substitution of the expression for $\vec{\none}$ (\ref{none}) yields
\begin{eqnarray}
\ntwo^{\mu} n_{\mu} = - \frac{1}{2} \Kperpernornor + \frac{3}{4} \sigma \Kpernornor^2 + 
\left ( \Kpernortan_{\mu} - \Qone a_{\mu} - \sigma D_{\mu} \Qone \right )
\left ( \Kpernortan^{\mu} +
\Qone a^{\mu} + \sigma D^{\mu} \Qone \right ). 
\label{n2n}
\end{eqnarray}
For the tangential components, it is convenient to use the second variation of the
normal one-form, i.e
$\bm{\mtwo} \equiv \partial_{\ee} \partial_{\ee} \bm{n}_{\ee} |_{\ee=0}$.
The relationship with $\vec{\ntwo}$ is immediate from the second $\ee$-derivative of
$g_{\ee} (\vec{n}_{\ee}, \cdot )= \bm{n}_{\ee}$, i.e.
\begin{eqnarray*}
\ntwo^{\alpha} = - \Kperper^{\alpha\beta} n_{\beta} + \sigma \Kpernornor \Kper^{\alpha\beta} n_{\beta} +
 2 \Kper^{\alpha\beta}
\left ( \Kpernortan_{\beta} + \Qone a_{\beta} + \sigma_{\beta} \Qone \right ) + \mtwo^{\alpha}.
\end{eqnarray*}
Decomposing this into tangential and normal components and using (\ref{n2n}) one finds
\begin{eqnarray}
\ntwo^{\alpha} = 
n^{\alpha} \left [ - \frac{1}{2} \sigma \Kperpernornor + \frac{3}{4} \Kpernornor^2 + \sigma 
\left ( \Kpernortan_{\mu} - \Qone a_{\mu} - \sigma D_{\mu} \Qone \right ) 
\left ( \Kpernortan^{\mu} +  \Qone a_{\mu} + \sigma D_{\mu} \Qone \right )  \right ] \nonumber \\
- \Kperpernortan^{\alpha} 
+ \sigma \Kpernornor \Kpernortan^{\alpha}
+ 2 \Kpertan^{\alpha\beta} \left ( \Kpernortan_{\beta} + \Qone a_{\beta} + \sigma D_{\beta} \Qone \right ) 
+ h^{\alpha}_{\beta} \mtwo^{\beta}.  \label{n2.1}
\end{eqnarray}
It only remains to find $h^{\alpha}_{\beta} \mtwo^{\beta}$. 
Applying Lemma \ref{FirstSecondOrder} to
$\Phi_{\ee} (\bm{n}_{\ee}) = 0$, $\forall \ee$ yields
\begin{eqnarray}
\Phi^{\star} \left ( \pounds_{\vec{\W}} \bm{n} + \pounds_{\vec{\Zone}} \pounds_{\vec{\Zone}} \bm{n} +
 2 \pounds_{\vec{\Zone}}
\bm{\mone} + \bm{\mtwo} \right ) = 0. \label{masterm2}
\end{eqnarray}
Moreover, Lemma  \ref{LieZZ} applied to $\bm{n}$ leads to
\begin{eqnarray*}
\pounds_{\vec{\W}} \bm{n} + \pounds_{\vec{\Zone}} \pounds_{\vec{\Zone}} \bm{n} = \pounds_{\vec{Z}_2} \bm{n}
+ \pounds_{\vec{\Tone}} \pounds_{\vec{\Tone}} \bm{n} + 2 \pounds_{\vec{\Tone}} \pounds_{\Qone\vec{n}} \bm{n}
- \pounds_{\Cone \vec{n} + \vec{\Done} } \bm{n} - \\
- \pounds_{\Qone^2 \vec{a}}  \bm{n} 
+ \pounds_{\Qone \vec{n}} \pounds_{\Qone \vec{n}} \bm{n}.
\end{eqnarray*}
Now, for any pair of functions $F_1$ and $F_2$ we have
$\pounds_{F_1 \vec{n}} \left ( F_2 \bm{n} \right )_{\alpha} =  \vec{n} ( F_1 F_2) n_{\alpha}
+ F_2  ( F_1 a_{\alpha} + \sigma D_{\alpha} F_1 )$. 
Using also the fact that $\Phi^{\star} ( \pounds_{\vec{V}} \bm{n}  ) = 0$ for any vector field $\vec{V}$
 tangent to $\Sigma$, i.e.
$\pounds_{\vec{V}} \bm{n} \propto \bm{n}$ one finds
\begin{eqnarray*}
\pounds_{\vec{\W}} n_{\alpha} + \pounds_{\vec{\Zone}} \pounds_{\vec{\Zone}} n_{\alpha}   = 
\Qtwo a_{\alpha} + \sigma D_{\alpha} \Qtwo + 
2 \pounds_{\vec{\Tone}} \left ( \Qone a_{\alpha} + \sigma D_{\alpha} \Qone \right ) - \left ( \Cone a_{\alpha} + \sigma
D_{\alpha} \Cone \right ) + \\
\vec{n} (\Qone) \left ( \Qone a_{\alpha} + \sigma D_{\alpha} \Qone \right ) +
 \pounds_{\Qone \vec{n}} \left ( \Qone a_{\alpha} + \sigma D_{\alpha} \Qone \right )  + P n_{\alpha}.
\end{eqnarray*}
Moreover, from (\ref{mone})
\begin{eqnarray*}
 \pounds_{\vec{\Zone}} \mone_{\alpha}  = - 
\pounds_{\vec{\Tone}} \left ( \Qone a_{\alpha} + \sigma D_{\alpha} \Qone \right )
+ \frac{\sigma}{2} \Kpernornor \left ( \Qone a_{\alpha} + \sigma D_{\alpha} \Qone \right ) 
- \pounds_{\Qone \vec{n}} \left ( \Qone a_{\alpha} + \sigma D_{\alpha} \Qone \right ) + P n_{\alpha}.
\end{eqnarray*}
Then (\ref{masterm2}) implies
\begin{eqnarray}
h_{\alpha}^{\beta} \mtwo_{\beta} = 
- \left ( \Qtwo a_{\alpha} + \sigma D_{\alpha} \Qtwo \right ) + \left ( \Cone a_{\alpha} + 
\sigma D_{\alpha} \Cone \right ) 
- \left ( \vec{n} (\Qone) + \sigma \Kpernornor \right ) \left ( \Qone a_{\alpha} + 
\sigma D_{\alpha} \Qone \right )+ \nonumber
\\ 
+ h^{\beta}_{\alpha}  \pounds_{\Qone \vec{n}} \left ( \Qone a_{\beta} + \sigma D_{\beta} \Qone \right ). 
\label{m2.1}
\end{eqnarray}
Let us elaborate the last term. Using $\pounds_{\Qone \vec{n}} h^{\alpha\beta}
= - 2 \Qone \kappa^{\alpha\beta} - \sigma n^{\alpha} ( \Qone a^{\beta} + \sigma D^{\beta} \Qone )
- \sigma n^{\beta} ( \Qone a^{\alpha} + \sigma D^{\alpha} \Qone )$, and integrating by parts
we find
\begin{eqnarray*}
h^{\alpha\beta} \pounds_{\Qone \vec{n}} \left ( \Qone a_{\beta} + \sigma D_{\beta} \Qone \right) 
= \left [ \Qone \vec{n}, \Qone \vec{a} + \sigma \gradL  \Qone \right ]^{\alpha}
+ 2 \Qone^2 a^{\mu} \kappa^{\alpha}_{\, \, \mu} + 2 \sigma \Qone 
\kappa^{\alpha}_{\, \, \mu} D^{\mu} \Qone \\
+ \sigma n^{\alpha} 
\left ( \Qone a_{\mu} + \sigma D_{\mu} \Qone \right )
\left ( \Qone a^{\mu} + \sigma D^{\mu} \Qone \right ).
\end{eqnarray*}
Plugging  this into (\ref{m2.1}) and the resulting expression in (\ref{n2.1}), the Lemma follows. $\hfill \Box$.


\begin{thebibliography}{999}

\bibitem{Carbone05} 
\Journal{C. Carbone, S. Matarrese}{Unified treatment of cosmological perturbations
from superhorizon to small scales}{\PRD}{2005}{\bf 71}{043508}

\bibitem{Kolb05}
\Journal{E.W. Kolb, S. Matarrese, A. Notari, A. Riotto}{Effect of
inhomogeneities on the expansion rate of the universe}{\PRD}{2005}{71}{023524}

\bibitem{Malik04}
\Journal{K.A. Malik, D. Wands}{Evolution of second-order cosmological perturbations} 
{\CQG}{2004}{21}{L65-L71}

\bibitem{Mena02}

\Journal{F.C. Mena, R. Tavakol, M. Bruni}{Second order perturbations of
flat dust FLRW universes with a cosmological constant}{\IJMPA}{2002}{17}{4239-4244} 


\bibitem{Matarresse98}
\Journal{S. Matarrese, S. Mollerach, M. Bruni}{Relativistic second-order
perturbations of the Einstein-de Sitter universe}{\PRD}{1998}{58}{043504}



\bibitem{Singh02}
\Journal{S. Singh, C.P. Ma}{Linear and second-order evolution of cosmic baryon perturbations below $10^6$
solar masses}{ \AJ}{2002}{569}{1-7}


\bibitem{Watts01} 
\Journal{P.I.R. Watts, A.N. Taylor}{Evolution of the probability distribution
function of galaxies in redshift space}{\MNRAS}{2001}{320}{139-152}


\bibitem{Bartolo04.1} 
\Journal{N. Bartolo, S. Matarrese, A. Riotto}{Gauge-invariant temperature anisotropies and primordial
non-Gaussianity}{\PRL}{2004}{93}{231301}

\bibitem{Bartolo04.2} 
\Journal{N. Bartolo, E. Komatsu, S. Matarrese, A. Riotto}
{Non-Gaussianity from inflation: theory and observations}
{\it Physics Reports}{2004}{402}{103-266}

\bibitem{Acquaviva03}
\Journal{V. Acquaviva, N. Bartolo, S. Matarrese, A. Riotto}
{Second-Order Cosmological Perturbations from Inflation}
{\NPB}{2003}{667}{119-148}

\bibitem{Hartle67}
\Journal{J. Hartle}{Slowly rotating relativistic stars I. Equations of
                  structure}{\AJ}{1967}{150}{1005-1029}

\bibitem{Stergioulas}
\Journal{N. Stergioulas}{Rotating Stars in Relativity}{\LRR}{2003}{6}{3} 
 http://www.livingreviews.org/lrr-2003-3.


\bibitem{Gleiser96}
\Journal{R.J. Gleiser, C.O. Nicasio, R.H. Price, J. Pullin}{Colliding black
holes: How far can the close approximation go?}{\PRL}{1996}{77}{4483-4486} 

\bibitem{Gleiser00}
\Journal{R.J. Gleiser, C.O. Nicasio, R.H. Price, J. Pullin}{Gravitational radiation
from Schwarzschild black holes: the second-order perturbation formalism}{\it Physics Reports}{2000}
{325}{42-81}

\bibitem{Garriga93} \Journal{J. Garriga, A. Vilenkin}{Black-holes from nucleating strings}{ 
\PRD}{1993}{47}{3265-3274}


\bibitem{Carter93} \Journal{B. Carter}{Perturbation dynamics for membranes and
strings governed by the Dirac-Goto-Nambu action in curved space}{\PRD}{1993}{48}{4835-4838}


\bibitem{Guven93} \Journal{J. Guven}
{Covariant perturbations of domain-walls in curved spacetime}{\PRD}{1993}{48}{4604-4608}


\bibitem{Battye95} \Journal{R.A. Battye, B. Carter}{Gravitational perturbations of relativistic membranes and strings}
{\PLB}{1995}{357}{29-35}

\bibitem{reviews} R. Maartens, ``Geometry and dynamics of the brane-world'', gr-qc/0101059,
in {\it Frames and Gravitomagnetism} Ed. J. Pascual-S\'anchez {\it et. al} (World Scientific, 2001), p.93-119;
D. Langlois, ``Brane cosmology: an introduction'', {\it Prog. Theor. Phys. Suppl.} {\bf 148}, 181-212 (2003);
N. Derruele, ``Linearized gravity on branes: from Newton's law to cosmological perturbations'',
in {\it Gravitation and Cosmology} Ed. A. Lobo {\it et al.} (Pub. Universitat de Barcelona, 2003), p.25-36.

\bibitem{MalikRodriguez-MartinezLanglois}
\Journal {K.A. Malik, M. Rodr\'{\i}guez-Mart\'{\i}nez, D. Langlois}
  	 {Defining perturbations on submanifolds} {\PRD} {2003} {68}
  	 {123517}

\bibitem{Maartens02}
\Journal{R. Maartens}{Brane-world cosmological perturbations - A
   	 covariant approach -}{\PTPS} {2002} {148} {213-234}


\bibitem{Mukohyama00}
\Journal {S. Mukohyama} {Perturbation of junction condition and doubly
  	 gauge-invariant variables} {\CQG} {2000} {17} {4777-4798}



\bibitem{KudohTanaka01}
\Journal{H. Kudoh, T. Tanaka}{Second order perturbations in the Randall-Sundrum
infinite brane-world model}{\PRD}{2001}{64}{084022}

\bibitem{Kudoh02}
\Journal{H. Kudoh}{Gravity beyond linear perturbations in the
braneworld}{\PTPS}{2002}{148}{145-157}

\bibitem{KudohTanaka02}
\Journal{H. Kudoh, T. Tanaka}{Second order perturbations in the radius stabilized
Randall-Sundrum two branes model}{\PRD}{2002}{65}{104034}

\bibitem{KudohTanaka03}
\Journal{H. Kudoh, T. Tanaka}{Second order perturbations in the radius stabilized
Randall-Sundrum two branes model. II. Effect of relaxing strong coupling approximation}
{\PRD}{2003}{67}{044011}

\bibitem{Battye00} \Journal{R.A. Battye, B. Carter}{Second-order Lagrangian and
symplectic current for gravitationally perturbed Dirac-Goto-Nambu strings and branes}{\CQG}{2000}
{\bf 17}{3325-3334}


\bibitem{Bruni97}
\Journal{M. Bruni, S. Matarrese, S. Mollerach, S. Sonego}{Perturbations of spacetime: gauge transformations
and gauge invariance at second order and beyond}{\CQG}{1997}{14}{2585-2606}


\bibitem{Sonego98}
\Journal{S. Sonego, M. Bruni}{Gauge Dependence in the Theory of Non-Linear
Spacetime Perturbations}{\CMP}{1998}{193}{209-218}

\bibitem{MarsSenovilla93}
\Journal {M. Mars, J.M.M. Senovilla} {Geometry of general
  	 hypersurfaces in spacetime - junction conditions} {\CQG}
  	 {1993} {10} {1865-1897}


\bibitem{Israel} W. Israel, ``Singular hypersurfaces
and thin shells in General Relativity'' {\it
Nuovo Cimento B} {\bf 44} 1 (1966); Correction {\bf 48} 463 (1967).

\bibitem{ES} A. Einstein, E.G. Straus, 
``The influence of the expansion of space on the gravitation fields surrounding the individual 
stars''
 {\it Rev. Mod. Phys.} {\bf 17},
120-124 (1945); erratum {\it ibid} {\bf 18} 148 (1946).

\bibitem{Malcolm}
M.A.H. MacCallum, M. Mars, R. Vera,
``First and second order exterior metrics of stationary and axially
symmetric slowly rotating stars'', in preparation.

\bibitem{Wald} R. Wald, "General Relativity", University of Chicago Press, Chicago, (1984).

\bibitem{ClarkeDray87}
\Journal{C.J.S. Clarke, T. Dray} {Junction conditions for null
  	hypersurfaces} {\CQG} {1987} {4} {265-275}

\bibitem{Kind93}
\Journal{S. Kind, J. Ehlers, B.G. Schmidt}
{Relativistic Stellar Oscillations Treated as an Initial-Value Problem}{\CQG}{1993}{10}{2137-2152}

\bibitem{Gerlach79.1}
\Journal{U.H. Gerlach, U.K. Sengupta}{Even parity junction conditions for perturbations on most general
spherically symmetric space-times}{\PRD}{1979}{20}{2540-2546}

\bibitem{Gerlach79.2}
\Journal{U.H. Gerlach, U.K. Sengupta}{Junction conditions for odd-parity perturbations on most general
spherically symmetric space-times}{\PRD}{1979}{20}{3009-3014}

\bibitem{MartinGarcia}
\Journal{J.M. Mart\'{\i}n-Garc\'{\i}a, C. Gundlach}{Gauge invariant and coordinate-independent perturbations
of stellar collapse II: matching to the exterior}{\PRD}{2001}{64}{024012}





\end{thebibliography}
\end{document}